\documentclass{aa}

\usepackage{graphicx}
\usepackage{txfonts}
\usepackage{hyperref}
\usepackage{multirow}
\usepackage[capitalise]{cleveref}
\usepackage[per-mode=symbol]{siunitx}
\usepackage{graphicx}
\usepackage{textcomp}
\bibpunct{(}{)}{;}{a}{}{,}

\DeclareSIUnit{\astronomicalunit}{au}

\newcommand{\e}[1]{\times 10^{#1}}
\newcommand{\md}{\text{d}}
\newcommand{\nut}{{\tilde \nu}}

\begin{document}

\title{Accuracy tests of radiation schemes used in hot Jupiter global circulation models}

\titlerunning{Accuracy tests of radiation schemes used in hot Jupiter global circulation models}

\author{%
David~Sk{\aa}lid~Amundsen\inst{1}\and%
Isabelle~Baraffe\inst{1}\and%
Pascal~Tremblin\inst{1}\and%
James~Manners\inst{2}\and%
Wolfgang~Hayek\inst{1}\and%
N.~J.~Mayne\inst{1}\and%
David~M.~Acreman\inst{1}%
}
\authorrunning{Amundsen  et al.}

\institute{%
Astrophysics Group, University of Exeter, Exeter, EX4 4QL, United Kingdom\\
\email{D.S.Amundsen@exeter.ac.uk}
\and
Met Office, Exeter, EX1 3PB, United Kingdom
}

\date{}

 

\abstract{
The treatment of radiation transport in global circulation models (GCMs) is crucial for correctly describing Earth and exoplanet atmospheric dynamics processes. The two-stream approximation and correlated-$k$ method are currently state-of-the-art approximations applied in both Earth and hot Jupiter GCM radiation schemes to facilitate the rapid calculation of fluxes and heating rates. Their accuracy have been tested extensively for Earth-like conditions, but verification of the methods' applicability to hot Jupiter-like conditions is lacking in the literature. We are adapting the UK Met Office GCM, the Unified Model (UM), for the study of hot Jupiters, and present in this work the adaptation of the Edwards--Slingo radiation scheme based on the two-stream approximation and the correlated-$k$ method. We discuss the calculation of absorption coefficients from high-temperature line lists and highlight the large uncertainty in the pressure-broadened line widths. We compare fluxes and heating rates obtained with our adapted scheme to more accurate discrete ordinate (DO) line-by-line (LbL) calculations ignoring scattering effects. We find that, in most cases, errors stay below $\SI{10}{\percent}$ for both heating rates and fluxes using $\sim 10$ $k$-coefficients in each band and a diffusivity factor $D = 1.66$. The two-stream approximation and the correlated-$k$ method both contribute non-negligibly to the total error. We also find that using band-averaged absorption coefficients, which have previously been used in radiative-hydrodynamical simulations of a hot Jupiter, may yield errors of $\sim \SI{100}{\percent}$, and should thus be used with caution.}


\maketitle

\section{Introduction}

For Earth's atmosphere, irradiation from the Sun is the primary source of energy. Any model of the Earth's atmosphere therefore needs a robust and accurate treatment of radiation transport. Global circulation models (GCMs) of the Earth are used for both weather prediction and climate research, and they include a dynamical core that solves some variant of the Navier--Stokes equations and a radiation scheme that calculates the radiative heating rate. The dynamical cores are tested using benchmarks \cite[see e.g.][]{Held1994,Reed2011,Ullrich2013}, and both dynamical cores\footnote{See \url{http://earthsystemcog.org/projects/dcmip-2012/}} and radiation schemes \citep{Ellingson1991,Collins2006,Oreopoulos2012} are tested through intercomparison projects.

GCMs have also been successfully applied to other solar system planets such as Jupiter, Saturn, Mars and Venus \citep[see for example][respectively]{Yamazaki2004, Muller-Wodarg2006, Hollingsworth2010, Lebonnois2011}. In the past decade, GCMs have been used to study large-scale circulation on hot Jupiters \citep{Showman2002,Showman2009,Rauscher2012,Cho2008,Thrastarson2010,Dobbs-Dixon2008}, a class of extrasolar planets which are approximately the size of Jupiter but orbit less than $\SI{0.1}{\astronomicalunit}$ from their parent star. These planets, thought to have tidally locked circular orbits due to strong tidal interactions between the planet and its parent star \citep{Baraffe2010}, experience intense irradiation yielding a significant temperature contrast between the (permanent) day-side and night-side. Winds in the atmosphere of these planets are therefore expected to transport heat from the day-side to the night-side.

Atmospheric properties of hot Jupiters are obtained by various observational techniques, mainly transmission spectroscopy and secondary eclipse measurements. Brightness maps \citep{Knutson2007,Majeau2012} and wind velocities \citep{Snellen2010} are now accessible, and constraints on the composition are becoming available, but large uncertainties remain. Observations indicate a hotspot shifted eastward of the substellar point \citep{Knutson2007,Majeau2012} and temperature contrasts smaller than what is expected for these planets without winds \citep{Knutson2007,Knutson2009}, indicating transport of heat from the day-side to the night-side \citep{Watkins2010,Perez-Becker2013}. HD~209458b appears to have a temperature inversion in its upper atmosphere \citep{Knutson2008,Burrows2007} while HD~189733b does not \citep{Charbonneau2008,Barman2008,Knutson2009}, indicating that, despite similar orbital properties, hot Jupiters may have very different circulation patterns that still need to be understood. GCMs are therefore very valuable when trying to understand the increasing amount of observations of these systems.

Benchmarking of the dynamical cores of GCMs applied to hot Jupiters has been used to investigate stability of the codes and discrepancies between them \citep{Heng2011,Menou2009,Bending2013,Mayne2013a,Mayne2013b,Polichtchouk2014}. These benchmarks, and early GCMs applied to hot Jupiters, used simple, parametrised radiation schemes termed ``Newtonian cooling'' or ``temperature forcing'', where the temperature is relaxed towards assumed equilibrium pressure--temperature ($P$--$T$) profiles on a given timescale \citep{Showman2002}. Pressure--temperature profiles and timescales can be estimated using one-dimensional time-dependent radiative transfer calculations \citep{Iro2005}, though such an approach has flaws: (i) The equilibrium $P$--$T$ profiles used in the forcing may have a limited accuracy, (ii) radiative timescales may also have a limited accuracy and will vary in a non-trivial way as a function of latitude, longitude and depth, (iii) the forcing parametrisation itself may not be physically realistic, though the use of time-averaged equilibrium states when analysing model results may make this less of an issue, (iv) the model flexibility is poor since for each new planet modelled, the forcing must be changed.

Later studies used more complicated schemes such as flux-limited diffusion \citep{Dobbs-Dixon2008} and the two-stream approximation \citep{Showman2009,Rauscher2012,Dobbs-Dixon2013}. For the opacity treatment, grey schemes \citep{Rauscher2012}, binning and averaging of the absorption coefficients \citep{Dobbs-Dixon2013} and the correlated-$k$ method \citep{Showman2009} have been used. The correlated-$k$ method has also been used for retrieval analysis and characterisation of hot Jupiter atmospheres \citep{Irwin2008} and to model brown dwarf atmospheres \citep{Burrows1997}. Brown dwarfs atmospheres have many similarities with hot Jupiter atmospheres (e.g. temperature range and composition), but local conditions are very different due to the strong irradiation from the parent stars for hot Jupiters. There is a notable lack of analysis of the accuracy of these schemes when applied to hot Jupiter-like atmospheres and of details on how opacities have been calculated from line lists, preventing rigorous comparison with results previously published in the literature. These are serious shortcomings in a field of research which develops quickly and will deliver more and more accurate data requiring reliable tools for their interpretation. 

Both the two-stream approximation and the correlated-$k$ method \citep[see e.g.][]{Thomas2002} are widely used in GCM simulations of the Earth, and the literature on the methods' applicability to the Earth atmosphere and their accuracy is extensive \citep[see e.g.][]{Toon1989,Meador1980,Zdunkowski1980,Goody1989,Lacis1991,Mlawer1997}. They have both been found to yield results with satisfactory accuracy when comparing to more accurate solutions obtained from e.g. discrete ordinate (DO), line-by-line (LbL) calculations and when different schemes are compared through intercomparison projects \citep{Ellingson1991,Collins2006,Oreopoulos2012}. They are, however, still under investigation \citep{Goldblatt2009} and are still one of the limiting factors of the accuracy of both weather prediction and climate modelling.

We are currently adapting the UK Met Office GCM, the Unified Model (UM), to hot Jupiter-like conditions. The main advantage of this model is its dynamical core, which solves the full 3D Euler equations and is coupled to a radiation scheme based on the correlated-$k$ method and two-stream approximation. This GCM is state-of-the-art for both Earth and hot Jupiter atmospheric dynamics modelling. In previous papers we have tested and confirmed the dynamical core's suitability to model hot Jupiter-like atmospheres \citep{Mayne2013a,Mayne2013b}. This paper presents the adaptation of the radiation scheme, the Edwards--Slingo (ES) radiation scheme \citep{Edwards1996a}, to conditions prevailing in hot Jupiter atmospheres. Opacities from high-temperature line lists have been calculated for the dominant absorbers in these atmospheres and the radiation scheme, using $k$-coefficients calculated from these opacities, has been tested against more accurate DO LbL calculations.

Observations of absorbing and scattering species in hot Jupiter atmospheres have so far been limited to the detection of molecular absorbers \citep{Huitson2013,Wakeford2013,Tinetti2007}, with some observations suggesting Rayleigh/Mie scattering clouds \citep{Pont2008,Sing2013}. Due to the large uncertainties related to scatterers in hot Jupiter atmospheres and the complexity it adds to radiation transport, we limit the discussions in this paper to purely absorbing atmospheres and postpone the inclusion of scattering to a future work.

The motivation for the present work is the lack of accurate tests and analysis of these radiation schemes now widely used by the community. This work will help in the implementation of similar schemes in the future and provide some guidelines for further progress. In \cref{sec:opacities} we first discuss our opacity data including the high-temperature line lists we use and the calculation of cross-sections from these line lists by using estimates for the line widths. In \cref{sec:ES} we briefly summarise the implementation of the correlated-$k$ method and two-stream approximation in the UM and move on to testing this scheme for gradually more complicated hot Jupiter-like atmospheres in \cref{sec:testing}. These tests will be useful to other groups for comparison and benchmark purposes. Our conclusions follow in \cref{sec:conclusions}. Combination of the adapted dynamical core and radiation schemes is in progress and will be presented in a future work, which will result in a state-of-the-art GCM that can be applied to a variety of exoplanet atmospheres.

\section{Opacities} \label{sec:opacities}

In this section we present the calculation of mass absorption coefficients from high-temperature line lists. We first discuss our opacity data, including line lists and line broadening parameters, and then provide details of our numerical implementation.

\subsection{Opacity data}

The dominant absorbers in hot Jupiter atmospheres with solar metallicity are H$_2$O, CO, CH$_4$, NH$_3$, TiO, VO and H$_2$--H$_2$ and H$_2$--He Collision Induced Absorption (CIA) \citep{Burrows1999,Baraffe2010}. References for the line lists and partition functions we use are given in \cref{tbl:opacity_data}. Where available we used line lists from the ExoMol project \citep{Tennyson2012} (H$_2$O, named BT2, and NH$_3$, named BYTe), while HITEMP \citep{Rothman2010} (CO) and the CIA extension to HITRAN \citep{Richard2012} (H$_2$--H$_2$ and H$_2$--He CIA) are also used. For line lists from the ExoMol project, partition functions are calculated explicitly by summing up the energy levels. For CO we use the HITRAN partition functions \citep{Rothman2009}, while for TiO and VO we use the polynomial fits in \citet{Sauval1984} as recommended by B. Plez (private communication). We use isotopic abundances, $I_a$, from \citet{Asplund2009}.

Our methane line list is calculated using the Spherical Top Data System (STDS) \citep{Wenger1998} setting the cut-off at $J=60$. This line list has several flaws, e.g. important bands in the absorption spectrum are missing. We are aware that a new ExoMol CH$_4$ line list should soon be released, but this will not change the main conclusions of this paper. We use the $^{12}$CH$_4$ partition function from \citet{Wenger2008}, and $^{13}$CH$_4$ partition function from HITRAN \citep{Rothman2009}.

\begin{table*}
\centering
\caption{List of molecules included in our opacity database with associated line list and partition function sources.}
\begin{tabular}{l|l|l}
Molecule & Line list & Partition function \\ \hline
H$_2$O & \cite{Barber2006} & \cite{Barber2006} \\ \hline
\multirow{2}{*}{CH$_4$} & \multirow{2}{*}{\cite{Wenger1998}} & \cite{Wenger2008} ($^{12}$CH$_4$) \\
 & & \cite{Rothman2009} ($^{13}$CH$_4$) \\
\hline
CO & \cite{Rothman2010} & \cite{Rothman2009} \\ \hline
NH$_3$ & \cite{Yurchenko2011} & \cite{Yurchenko2011} \\ \hline
TiO & \cite{Plez1998} & \cite{Sauval1984} \\ \hline
VO & B. Plez (private communication) & \cite{Sauval1984} \\ \hline
H$_2$--H$_2$ CIA & \cite{Richard2012} & N/A \\ \hline
H$_2$--He CIA & \cite{Richard2012} & N/A \\ \hline
\end{tabular}
\label{tbl:opacity_data}
\end{table*}

Due to the very large number of lines in these line lists, local thermodynamic equilibrium (LTE) is assumed to calculate level populations. Non-LTE effects could be important in the upper part of irradiated planet atmospheres. Indeed some observational works invoke non-LTE effects to explain the detection of strong emission features in hot Jupiters \citep[see e.g.][]{Waldmann2012}, and there has been some work on modelling non-LTE effects in these atmospheres \citep{Schweitzer2004,Barman2002}. The significance of these effects still needs to be proven, however, and it is beyond the scope of the present work to consider non-LTE effects.

The line intensity for transition $i$ can then be calculated by using \citep{Thomas2002}
\begin{align}
\mathcal S_i(T) &= \frac{I_a}{8\pi c \nut_0^2} \frac{g_\text{u} e^{-E_\text{l}/k_\text{B}T}}{Q(T)} \left( 1 - e^{- hc\nut_0/k_\text{B} T} \right) A_i
\label{eq:S_i_A} \\
&= \frac{I_a \pi e^2}{m_\text{e} c^2} \frac{e^{-E_\text{l}/k_\text{B}T}}{Q(T)} \left( 1 - e^{- hc\nut_0/k_\text{B} T} \right) g_\text{l} f_\text{lu},
\label{eq:S_i_gf}
\end{align}
where $T$ is the local temperature, $c$ is the velocity of light, $\nut_0$ is the wavenumber of the transition, $g_\text{u}$ is the degeneracy of the upper level, $E_\text{l}$ is the energy of the lower level, $k_\text{B}$ is Boltzmann's constant, $Q(T)$ is the partition function evaluated at $T$, $A_i$ is the Einstein $A$-coefficient of the transition, $e$ is the electron charge in CGS-Gaussian units, $m_\text{e}$ is the electron mass and $g_\text{l} f_\text{lu}$ is the $gf$-value of the transition with $g_\text{l}$ and $f_\text{lu}$ being the degeneracy of the lower level and the oscillator strength, respectively. The quantities $\nut_0$, $g_\text{u}$, $E_\text{l}$ and $A_i$, or alternatively $g_\text{l} f_\text{lu}$, are given in the line lists. For line lists in the HITRAN format, line intensities $S_i(T_0)$ evaluated at a reference temperature $T_0 = \SI{296}{\kelvin}$ are given, and can be converted to any other temperature by using
\begin{equation}
\mathcal S_i (T) = \mathcal S_i (T_0)
\frac{Q(T_0)}{Q(T)}
\frac{e^{-E_\text{l}/k_\text{B}T}}{e^{-E_\text{l}/k_\text{B}T_0}}
\frac{\left( 1 - e^{- hc\nut_0/k_\text{B} T} \right)}{\left( 1 - e^{- hc\nut_0/k_\text{B} T_0} \right)} ,
\end{equation}
which follows from \cref{eq:S_i_A} by taking $S_i(T)/S_i(T_0)$. Both ionisation and molecular dissociation are ignored.

We calculate absorption coefficients including both Doppler broadening and pressure broadening from collisions with H$_2$ and He, which are the dominant species in hot Jupiter atmospheres, and ignore other broadening processes such as natural, self- and turbulent broadening as they are relatively unimportant compared to the former in hot Jupiter atmospheres. The pressure broadened width $\alpha_\text{L}$ depends on both pressure and temperature in a complex way, but the relationship is often approximated as \citep{Thomas2002,Sharp2007}
\begin{equation}
\alpha_\text{L}^z (P_z,T) = \alpha_{\text{L}}^z(P_0,T_0) \left( \frac{T_0}{T} \right)^{n_z} \frac{P_z}{P_0},
\label{eq:alpha_L^z}
\end{equation}
where $T_0$ and $P_0$ are the reference pressure and temperature, respectively, and $z$ is the perturbing species with $P_z$ the partial pressure of species $z$. The total pressure broadened width is the sum of the H$_2$ and He broadened widths. Pressure broadening parameters are, however, not included in the line lists and must be estimated from other sources. \Cref{tbl:line_widths} lists our pressure broadened width data sources, which are mostly gathered from experimental data, and partly overlap with those used by \citet{Bailey2012}. In these sources, $\alpha_{\text{L}}^z(P_0,T_0)$ is listed as a function of quantum numbers at a given temperature and pressure. The temperature dependence exponent $n_z$ also depends on the quantum numbers of the transition, but less so than $\alpha_{\text{L}}^z(P_0,T_0)$. We use the same $n_z$ for all transitions for a given species and broadener, a mean of the values found in the relevant paper(s) if more than one value is given. This approach is similar to that used by \citet{Sharp2007}.

\begin{table*}
\centering
\caption{Overview of our line width sources for pressure-induced broadening by hydrogen and helium.}
\begin{tabular}{l|l|l|l}
Molecule & Broadener & $\alpha_{\text{L}}^z(P_0,T_0)$ source & $n_z$ source \\ \hline
\multirow{2}{*}{H$_2$O} & H$_2$ & \citet{Gamache1996} & \multirow{2}{*}{\citet{Gamache1996}} \\
 & He & \citet{Solodov2009}, \citet{Steyert2004} & \\ \hline
\multirow{2}{*}{CH$_4$} & H$_2$ & \citet{Pine1992}, \citet{Margolis1993} & \citet{Margolis1993} \\
 & He & \citet{Pine1992} & \citet{Varanasi1990} \\ \hline
\multirow{2}{*}{CO} & H$_2$ & \citet{Regalia-Jarlot2005} & \citet{LeMoal1986} \\
 & He & \citet{BelBruno1982}, \citet{Mantz2005} & \citet{Mantz2005} \\ \hline
\multirow{2}{*}{NH$_3$} & H$_2$ & \multirow{2}{*}{\citet{Hadded2001}, \citet{Pine1993}} & \citet{Nouri2004} \\
 & He & & \citet{Sharp2007} \\ \hline
\multirow{2}{*}{TiO} & H$_2$ & \multirow{2}{*}{\citet{Sharp2007}, Eq.~(15)} & \multirow{2}{*}{\citet{Sharp2007}} \\
 & He &  & \\ \hline
\multirow{2}{*}{VO} & H$_2$ & \multirow{2}{*}{\citet{Sharp2007}, Eq.~(15)} & \multirow{2}{*}{\citet{Sharp2007}} \\
 & He & &
\end{tabular}
\label{tbl:line_widths}
\end{table*}

\begin{figure}
\includegraphics[width=\columnwidth]{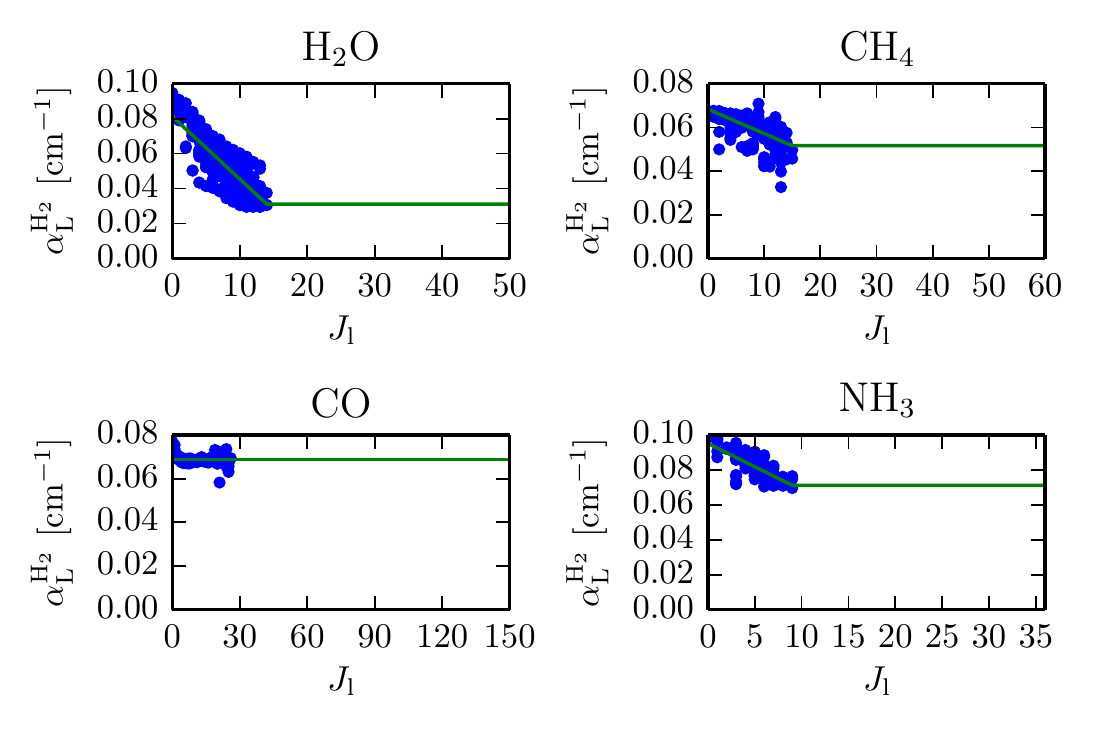}
\caption{H$_2$ pressure broadened line widths for H$_2$O, CH$_4$, CO and NH$_3$ from the sources listed in \cref{tbl:line_widths} at $P_\text{H$_2$} = \SI{e5}{\pascal}$ and $T = \SI{296}{\kelvin}$. Blue dots are the raw data points, the solid line is our fit and extrapolation. Conversion to other temperatures and pressures are done by using \cref{eq:alpha_L^z}.}
\label{fig:line_widths}
\end{figure}

In \cref{fig:line_widths}, we have plotted H$_2$ broadened line widths as a function of $J_\text{l}$, the total rotational quantum number of the lower level of the transition, for the sources listed in \cref{tbl:line_widths}. Since we have $\alpha_{\text{L}}^z(P_0,T_0)$ only for a small fraction of the transitions in the line lists, we model it as a linear function of $J_\text{l}$ using least squares fits. This is similar to previous line width studies \citep{Voronin2010,Burrows1997}.%
\footnote{Note that for CO the line width is approximately constant as a function of $J_\text{l}$, and we therefore use a simple mean instead of a linear fit, as shown in \cref{fig:line_widths}.} %
Unfortunately $\alpha_{\text{L}}^z(P_0,T_0)$ has to be extrapolated to high $J_\text{l}$ values, which we do by keeping it constant at the value where data for the highest $J_\text{l}$ is available to avoid introducing additional complexity (see \cref{fig:line_widths}). A different extrapolation scheme was tested, where widths were decreased further before flattening out. The effect on the local absorption coefficient was quite large, but averaging over many lines yields negligible difference between the two extrapolation schemes. Our final line widths as a function of $J_\text{l}$ are shown in \cref{fig:line_widths} as solid lines.

Note that $\alpha_{\text{L}}^z(P_0,T_0)$ and $n_z$ from the referenced papers have been obtained at approximately room temperature and pressure, while we are using \cref{eq:alpha_L^z} to extrapolate to both very high temperatures and pressures. Unfortunately there is currently no viable alternative. Any calculation of absorption coefficients including pressure broadening has to extrapolate the line width data, but we note that details on extrapolation schemes used in the literature is sparse.

Having obtained both line intensities and widths, the line profile cross section, $k_n^i(\nut)$, and mass absorption coefficient, $k_\rho^i(\nut)$, can be calculated using \citep{Thomas2002}:
\begin{align}
k_n^i(\nut) &= \mathcal S_i \frac{a}{\pi^{3/2} \alpha_\text{D}} \int_{-\infty}^\infty \frac{\md y \, e^{-y^2}}{(v - y)^2 + a^2} \equiv \mathcal S_i \Phi_\text{V}(\mathcal \nut), \\
k_\rho^i(\nut) &= \frac{k_n^i(\nut)}{\bar M_z}
\label{eq:Phi_V}
\end{align}
where $\bar M_z$ is the mean molecular weight of molecule $z$ in $\si{\kilogram \per molecule}$ and
\begin{equation}
v = (\nut - \nut_0)/\alpha_\text{D}, \quad
a = \frac{\alpha_\text{L}}{\alpha_\text{D}}, \quad
\alpha_\text{D} \equiv
\frac{\nut_0}{c} \sqrt{\frac{2k_\text{B}T}{\bar M_z}} .
\end{equation}
Here, $\alpha_\text{D}$ is the Doppler broadened width and $\Phi_\text{V}(\mathcal \nut)$ is the Voigt profile. The high velocity winds expected in the atmospheres of these planets will induce small Doppler shifts of the line centres, $\nut_0$. Non-zero pressure will also cause a shift of the line centres, but data on pressure-induced line-shifts is even more spare than that on line widths. The net effect of both Doppler and pressure-induced line shifts are uncertain, and we choose to ignore both.

We calculate mass absorption coefficients by summing the Voigt profiles,
\begin{equation}
k_\rho(\nut) = \sum_i k_\rho^i(\nut),
\end{equation}
on a fixed wavenumber grid using a grid spacing of $\SI{1e-3}{\centi \metre^{-1}}$.

CIA is a continuous opacity source, and is consequently provided in a different format. In the HITRAN format, $A$--$B$ CIA data is tabulated as the absorption per density of species $A$ per density of species $B$ as a function of temperature on a fixed wavenumber grid with $\SI{1}{\centi \metre^{-1}}$ spacing. Here, $A$ is always H$_2$, and we multiply the CIA data by the density of H$_2$ to make it consistent with the molecular line data.

\subsection{Numerical considerations}

The use of high-temperature line lists raises computational issues due to their size. In the HITRAN~2012 database \citep{Rothman2013}, the NH$_3$ line list has about $4.6\e{4}$ lines, while the ExoMol NH$_3$ line list, BYTe \citep{Yurchenko2011}, has about $1.1\e{9}$ lines. Consequently, calculating absorption coefficients from high-temperature line lists is significantly more computationally expensive than using smaller line lists such as HITRAN. In the literature this problem is often overcome by ignoring all lines with line intensities smaller than some value, sometimes evaluated at a fixed temperature \citep{Sharp2007}. The line intensity is, however, a strong function of temperature, and knowing where to apply the cut-off may be difficult.

A second cut-off has to be made, both for physical and computational reasons. According to \cref{eq:Phi_V} lines are infinite in extent and follow a Voigt profile. Unfortunately, real line profiles are not perfectly Voigtian \citep{Thomas2002}. The Voigt profile is fairly accurate provided interactions between molecules are weak, but for stronger interactions effects such as collisional narrowing may occur. Line wings are most affected, and to avoid overestimating the line wing absorption, it is common practice to apply a cut-off at some distance $d$ from the line centre \citep{Freedman2008,Sharp2007}. This distance may be fixed or be a function of pressure and/or temperature. In addition, evaluation of the Voigt profile is computationally expensive, and computing the line profiles to distances where it can be neglected adds an unnecessary computational cost.

To cope with these problems we have developed a scheme to combine the line wing cut-off with an elimination of unimportant weak lines to decrease computation time. The cut-off distance $d$ is calculated on-the-fly by estimating when the line mass absorption coefficient has reached some value, $k_\rho^{\text{cut}}$. This is done by approximating the line profile as Lorentzian with a width equal to the sum of the Doppler and pressure broadened widths to facilitate analytical treatment and ensure that the profile width is not underestimated. This yields the following formula for $d$:
\begin{equation}
d = \sqrt{\tilde \alpha \max\left( \frac{\mathcal S}{\pi k_\rho^\text{cut}} - \tilde \alpha , 0 \right)}, \qquad
\tilde \alpha = \alpha_\text{L} + \alpha_\text{D},
\end{equation}
For very weak lines, $\mathcal S_i/\pi k_\rho^\text{cut} - \tilde \alpha < 0$, i.e. the value of the line mass absorption coefficient at the line centre is smaller than $k_\rho^\text{cut}$, and consequently the line can be ignored completely.

Lines are added one-by-one to the total mass absorption coefficient spectrum. The value of $k_\rho^\text{cut}$ is chosen to be some fraction $f_\text{AK}$ of the latest value of the total mass absorption coefficient at the line centre as line profiles are summed up: $k_\rho^\text{cut} = f_\text{AK} k_\rho^\text{latest}(\nut_0)$. We use the abbreviation AK (adaptive $k_\rho^\text{cut}$) to denote this cut-off method. Some lines can become unrealistically broad, however, so we include an upper limit on $d$ of $\SI{100}{\centi \metre^{-1}}$. Note that this cut-off scheme cannot be used if the water continuum is included since it requires a cut-off at a fixed distance from the line centre \citep{Clough1989}.

The main motivation for using this scheme is computational efficiency and not physical considerations, and the final absorption coefficient will depend slightly on the order in which lines have been added up. The advantages are, however, that weak lines can be discarded on-the-fly taking into account the line intensity at the current temperature, making a simple cut-off in line intensity unnecessary. It also ensures that strong lines are computed to larger distances from the line centres than weaker lines. The current lack of robust line broadening schemes for conditions characteristic of hot Jupiters forces us to use such artificial schemes.

We have, however, tried to limit the impact of our treatments by testing other schemes used in the literature. We compare our line profile cut-off scheme to two other schemes: (i) A cut-off at a fixed distance $d_\text{FW}$ from the line centre (fixed width, FW) and (ii) a cut-off at a distance from the line centre given by the sum of the Doppler and pressure broadened widths multiplied by some factor $f_\text{FF}$ (fixed factor, FF). The former scheme is similar to that used when including the water continuum from \citet{Clough1989}, where all lines have to be cut-off $\SI{25}{\centi \metre^{-1}}$ from the line centre, while the latter is similar to that used by \citet{Sharp2007}. Note that when using FF, we still apply the upper limit of $\SI{100}{\centi \metre^{-1}}$ on $d$.

\begin{figure}
\includegraphics[width=\columnwidth]{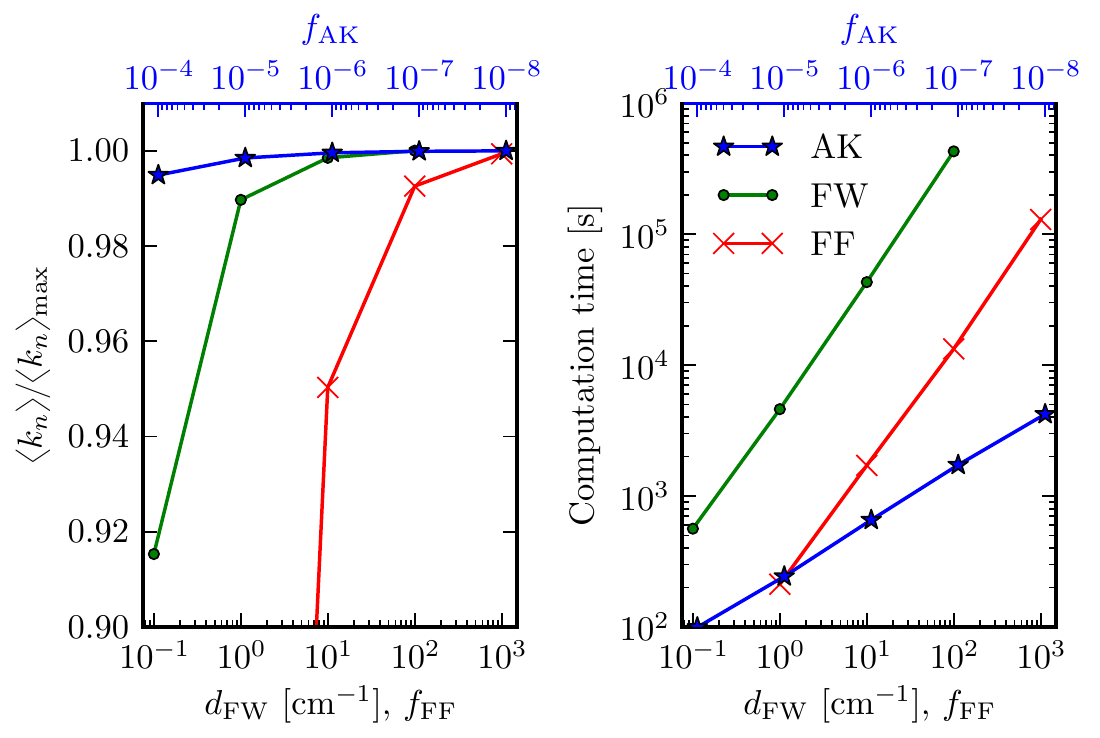}
\caption{Left-hand panel: Arithmetic mean of the H$_2$O absorption coefficient between $\SI{1000}{\centi \metre^{-1}}$ and $\SI{1001}{\centi \metre^{-1}}$ calculated using the adaptive (AK), fixed width (FW), fixed factor (FF) cut-off schemes as a function of the cut-off parameters ($f_\text{AK}$, $d_\text{FW}$ and $f_\text{FF}$, respectively) at $\SI{e5}{\pascal}$, $\SI{1500}{\kelvin}$. The mean absorption coefficients have been normalised by the value obtained using AK with $f_\text{AK} = \num{e-8}$. Right-hand panel: Computation time required using $12$ cores at $\SI{2.8}{\giga \hertz}$ as a function of the cut-off parameter. We see that our adaptive cut-off scheme is about two orders of magnitude faster than the two other methods for a given average absorption coefficient.}
\label{fig:line_wing_cutoff_scheme}
\end{figure}

In~\cref{fig:line_wing_cutoff_scheme}, we show both the average absorption coefficient between $\SI{1000}{\centi \metre^{-1}}$ and $\SI{1001}{\centi \metre^{-1}}$ at $\SI{e5}{\pascal}$, $\SI{1500}{\kelvin}$ and the computation time required for the three schemes as a function of the cut-off parameters. The code has been parallelised and runs using an Intel Xeon X5660 processor with $12$ cores at $\SI{2.8}{\giga \hertz}$.%
\footnote{For NH$_3$, having the largest of our adopted line lists with $>1.1$ billion lines, computation of absorption coefficients take about 11 days using our adaptive cut-off scheme on a computer with an Intel Xeon X5660 processor with $12$ cores at $\SI{2.8}{\giga \hertz}$.}%

The three cut-off methods reach approximately the same average absorption coefficient for the largest values of the cut-off distances (see \cref{fig:line_wing_cutoff_scheme}). Comparing the levels in the left-hand panels to the computation times in the right-hand panel, however, clearly shows the advantage of the AK method. At $f_\text{AK} = \num{e-6}$, this method reaches approximately the same level as FW at $d_\text{FW} = \SI{e2}{\centi \metre^{-1}}$ and FF at $f_\text{FF} = \num{e3}$. The computation time is, however, more than two orders of magnitude smaller. Due to the uncertainties in line widths and the significant decrease in computation time, we have decided to adopt this scheme for all our molecules using $f_\text{AK} = \num{e-6}$ except for CO.

CO lines are divided into several clearly separate, narrow bands. Consequently, the absorption coefficient will vary by many orders of magnitude on the scale of the bands. The AK scheme is unsuited to such situations since it will tend to produce large cut-off distances at the beginning of the cross-section calculation. These line wings will normally be hidden by stronger lines, but for CO they become non-negligible due to the lack of strong lines in certain wavelength regions. For CO we therefore use the FF method with $f_\text{FF} = \num{e2}$.

Absorption coefficients are tabulated on a $\log P, \log T$ grid with $30$ pressure points between $\SI{e-1}{\pascal}$ and $\SI{e8}{\pascal}$ and $20$ temperature points between $\SI{70}{\kelvin}$ and $\SI{3000}{\kelvin}$, uniformly distributed on a logarithmic scale.

\subsection{Molecular abundances} \label{sec:abundances}

For the present work and tests we use the closed-form expressions for the chemical equilibrium abundances of H$_2$O, CO, CH$_4$ and NH$_3$ from \citet{Burrows1999}. We use solar-like elemental abundances, listed in \cref{tbl:abundances}. We assume the gas to be ideal and H$_2$ and He partial pressures are calculated by assuming the atmosphere to be pure hydrogen and helium with an atomic hydrogen number fraction of $0.91183$. This yields a mean molecular weight of $\SI{2.3376}{\gram \per \mole}$. For a mixture of gases, $k$-coefficients are calculated from an effective mass absorption coefficient table obtained by summing mass absorption coefficients for each species weighted by the respective mass mixing ratios, similar to the approach in \citet{Showman2009}. Alternatively, the random overlap method can be used to combine $k$-coefficients for individual gases, though this is much more computationally expensive \citep{Lacis1991}. In \cref{sec:test1}, however, we only include absorption by H$_2$O and use $k$-coefficients calculated from the H$_2$O mass absorption coefficient and multiply the $k$-coefficients by the mass mixing ratio in each atmospheric layer.

\begin{table}
\centering
\caption{Our adopted solar-like elemental abundances \citep{Asplund2009}.}
\label{tbl:abundances}
\begin{tabular}{l|r}
Element (X) & $\log N_\text{X}/N_\text{H} + 12$ \\ \hline
C & $8.50$ \\
N & $7.86$ \\
O & $8.76$ \\
Si & $7.51$ \\
Ti & $4.95$ \\
V & $3.93$
\end{tabular}
\end{table}

For TiO and VO we use for the present tests a simple parametrisation scheme to prescribe their abundances. Ti and V are thought to sequester deeper than $\SI{e6}{\pascal}$ to $\SI{e7}{\pascal}$, while at low temperatures Ti and V will be bound to condensates \citep{Showman2009,Fortney2006}. We therefore parametrise the abundance of TiO and VO by assuming no absorbing TiO and VO to be present in the atmosphere for temperatures below $T_\text{crit}$ or pressure above $P_\text{crit}$. For temperatures above $T_\text{crit}$ and pressures below $P_\text{crit}$, however, we assume TiO and VO to be present, with partial pressures estimated by assuming all Ti bound in TiO and similarly all V bound in VO. This is a reasonable assumption since the abundances of both Ti and V are much smaller than that of oxygen, see \cref{tbl:abundances}. TiO and VO will therefore have a negligible effect on the availability of oxygen in the atmosphere. We use $T_\text{crit} = \SI{1500}{\kelvin}$, $P_\text{crit} = \SI{e7}{\pascal}$.

Non-equilibrium chemistry, due to photochemical or mixing processes, might be important in hot Jupiter atmospheres \citep{Cooper2006}. We are currently developing a non-equilibrium chemistry scheme, but for the sake of present tests we limit our analysis to equilibrium chemistry. Neither the exact form of the TiO and VO abundance nor the inclusion of non-equilibrium chemistry will, however, alter the main conclusions of this paper.

\section{The Edwards--Slingo radiation scheme and Atmo} \label{sec:ES}

Before discussing the tests we have performed to investigate the accuracy of the UM radiation scheme, we briefly summarise the formulation of the two-stream approximation \citep[see e.g.][]{Thomas2002} and correlated-$k$ method \citep{Lacis1991,Goody1989} implemented in the Edwards--Slingo (ES) radiation scheme \citep{Edwards1996a} currently used in the UM. This scheme has been widely used in the meteorological community, see e.g. \citet{Sun2011}. We also describe the line-by-line discrete ordinate code Atmo used for comparisons. Both models approximate the atmosphere as plane parallel, a standard approximation for radiation schemes in GCMs, but we note that sphericity my be important particularly near the terminator.


\subsection{The two-stream approximation}

The ES radiation scheme uses the two-stream approximated version of the radiative transfer equation as formulated by \citet{Zdunkowski1980,Zdunkowski1982,Zdunkowski1985} to obtain fluxes and heating rates, with details available in \citet{Edwards1996a}. Without scattering, the equations reduce to
\begin{equation}
\pm \frac{1}{D} \frac{\md F_{\nut,\text{d}}^\pm}{\md \tau (\nut)} = F_{\nut,\text{d}}^\pm - \pi B_\nut (T),
\label{eq:RT_TS}
\end{equation}
where $F_{\nut,\text{d}}^+$ and $F_{\nut,\text{d}}^-$ are the diffuse%
\footnote{Diffuse radiation is in general defined as thermal and scattered radiation. Since we ignore scattering, the diffuse radiation will simply be the thermal component.} %
upward and downward fluxes, respectively, $\tau(\nut)$ is the optical depth, $B_\nut (T)$ is the Planck intensity at temperature $T$ and $D$ is the diffusivity. Per definition, the downward diffuse flux is always zero at the upper boundary. The direct component%
\footnote{The direct component is defined as the unscattered part of the stellar radiation.} %
of the flux, $F_\text{s}^\pm$, is given by
\begin{align}
F_{\nut,\text{s}}^+ &= 0 &
F_{\nut,\text{s}}^- &= F_{\nut,\text{s}} e^{-\tau(\nut)/\mu_0}/\mu_0,
\end{align}
where $F_{\nut,\text{s}}$ is the solar flux at the top of the atmosphere and $\mu_0 = \cos \theta_0$ where $\theta_0$ is the solar zenith angle. The total upward and downward flux is the sum of the diffuse and direct components:
\begin{equation}
F_\nut^\pm = F_{\nut,\text{d}}^\pm + F_{\nut,\text{s}}^\pm, \\
\end{equation}
and we define the total flux, $F_\nut$, as the upward flux minus the downward flux, i.e.
\begin{equation}
F_\nut = F_\nut^+ - F_\nut^-.
\end{equation}
In the following, the explicit wavenumber dependence will be dropped to simplify the notation. Compared to the two-stream equation in \citet{Toon1989} this formulation is slightly different in that the thermal source function is $\pi B$ and not $2\pi B/D$. The choice $\pi B$ ensures the correct thermal source flux independent of the choice of $D$, and is consistent with the formulation previously used by \citet{Dobbs-Dixon2013,Rauscher2012}. Note that, for the thermal component, the ES radiation scheme solves the two-stream equations for the differential fluxes (total component of the flux less the Planck flux at the local temperature), see \citet{Edwards1996b}. It is also worth noting that \citet{Showman2009} used the two-stream source function method \citep{Toon1989} for the thermal component, which is exact in the no scattering case. We show that the two-stream approximation yields fairly accurate results for the thermal component without scattering, however, and plan to include scattering particles in the future. In the absence of scattering, the two-stream approximation yields the exact stellar component due to the separation of the flux into a direct (stellar) component and a diffuse (stellar) component, the latter being $0$ without scattering.%
\footnote{Note that for very hot planets, or planets with a low mass parent star (i.e. with a flux shifted towards the infrared), the thermal and stellar components will overlap significantly in wavelength. We specifically use the terminology \emph{thermal} and \emph{stellar components} over \emph{infrared} and \emph{visible wavelength regions} to avoid confusion.}

Since we ignore scattering, eliminating the single scattering albedo and backscattering coefficient, the only free parameter is the diffusivity factor $D$. It is related to the mean angle of the radiation by $\bar \mu = 1/D$, where $\bar \mu = |\cos \bar \theta|$ and $\bar \theta$ is the mean zenith angle. In the present work we explore three different values for $D$: $1.66$, originally due to \citet{Elasser1942}, $\sqrt{3} \approx 1.73$ from the discrete ordinate method with two quadrature points, and $2$, obtained assuming an isotropic radiation field \citep{Thomas2002}. See discussions in e.g. \citet{Thomas2002,Edwards1996b} for more details.

The optical depth is given by
\begin{equation}
\md \tau = -k(z) \, \md z = -k_\rho(z) \rho(z) \, \md z = -\sum_i \zeta_i(z) k_\rho^i (z) \rho(z) \, \md z,
\label{eq:dtau}
\end{equation}
or in integrated form, assuming hydrostatic equilibrium:
\begin{align}
\tau &= \sum_i \tau_i = \sum_i \int_z^\infty \md z' \, \zeta_i (z') k_\rho^i (z') \rho(z') \\
&= \frac{1}{g} \sum_i \int_0^P \md P' \, \zeta_i (P') k_\rho^i (P'),
\label{eq:tau}
\end{align}
where $k_\rho^i$ and $\zeta_i$ are the mass absorption coefficient and mass mixing ratio of species $i$, respectively. The total mass absorption coefficient is $k_\rho (z)$, and the total absorption coefficient is $k(z) = k_\rho(z) \rho(z)$.

The heating rate is given by
\begin{equation}
\mathcal H = -\frac{\md F}{\md z} = \frac{g \bar m P}{R T} \frac{\md F}{\md P}
\label{eq:heating_rate}
\end{equation}
where $g$ is the gravitational acceleration, $\bar m$ the mean molecular weight in $\si{\gram \per \mole}$, $P$ the pressure, $R$ the ideal gas constant and $F$ the total flux integrated over the entire spectrum. Hydrostatic equilibrium is assumed and the ideal gas equation is used to derive the final expression.

\subsection{The correlated-\texorpdfstring{$k$}{k} method}

Currently the ES radiation scheme uses a combination of the exponential sum fitting of transmissions (ESFT) technique \citep{Wiscombe1977} and the correlated-$k$ method \citep{Lacis1991,Goody1989} to obtain $k$-coefficients, some details of which can be found in \cite{Sun2011}. In each band the absorption coefficients from the line-by-line wavenumber grid are reordered according to strength in increasing order and divided into $n_k$ subintervals. The spacing of these subintervals must be the same for each $P$--$T$ and would ideally be spaced logarithmically in $k$.  Rather than using logarithmic $k$-intervals defined at a particular $P$--$T$ point, an average absorption coefficient $k_{\rho,\text{avg}}(\nut)$ is calculated from the top of the atmosphere down to an optical depth of one:
\begin{align}
k_{\rho,\text{avg}}^i (\nut) u_{\tau=1}^i &= \int_{z_{\tau=1}}^{\infty} \md z' \, \zeta_i (z') \rho(z') k_\rho^i (\nut, z') \\
&= \frac{1}{g} \int_0^{P_{\tau=1}} \md P' \, \zeta_i (P') k_\rho^i(\nu, P') \equiv 1,
\label{eq:k_avg_tau=1}
\end{align}
where $\zeta_i (z)$ is the mass mixing ratio of species $i$, $u_{\tau=1}^i$ is the column density of species $i$ down to $\tau = 1$ and hydrostatic equilibrium has been assumed. This is similar to the approach of \citet{Hogan2010} and provides an optimal sorting for the $P$--$T$ where each part of the spectrum is most important. We use an isothermal $P$--$T$ profile at $\SI{1116}{\kelvin}$, one of the temperatures in our $P$--$T$ grid, for this calculation as a compromise between day-side and night-side $P$--$T$ profiles of hot Jupiters. These average absorption coefficients are then used for the initial calculation of $k$-coefficients. In the following, the species index $i$ will be dropped for ease of notation.

The $k$-coefficient for subinterval $l$ is found by fitting the transmission, $e^{k_{\rho,\text{opt}}^l u_j}$, to the weighted transmissions of the line-by-line coefficients over a set of $n_u$ column densities, $u_j$, i.e.
\begin{equation}
\int_{g_l}^{g_{l+1}} \md g \, w(g) e^{k_\rho(g) u_j} \approx e^{k_{\rho,\text{opt}}^l u_j} , 
\end{equation}
where $w$ is the weighting function and $k_{\rho,\text{opt}}^{l}$ is the optimal $k$-coefficient in subinterval $l$. The probability of finding an absorption coefficient between $k$ and $k + \md k$ is $\md g = f(k) \, \md k$, with $g(k)$ being the cumulative probability distribution. In the above equation, $g_l$ is the $g$-coordinate corresponding to the beginning of the subinterval for $k$-term $l$ and $g_{n_k+1}$ is the $g$-coordinate for the end of $k$-term $n_k$. The error for $k$-term $l$ is defined as the root mean square (RMS) of the difference between the fitted exponential and exact integral for all column densities:
\begin{equation}
\epsilon_l = \sqrt{\frac{1}{n_u} \sum_{j=1}^{n_u} \left( \int_{g_l}^{g_{l+1}} \md g \, w(g) e^{k_\rho (g) u_j} - e^{k_{\rho,\text{opt}}^l u_j} \right)^2} ,
\end{equation}
The total error in a band, $\epsilon$, is defined as
\begin{align}
\epsilon &= \sqrt{\sum_{l=1}^{n_k} w_l \epsilon_l^2}, & \text{with} & &
w_l &= \int_{g_l}^{g_{l+1}} \md g \, w(g) .
\end{align}
The weights are defined to be normalised over each band:
\begin{equation}
\int_0^1 \md g \, w(g) = \sum_{l=1}^{n_k} w_l = 1,
\end{equation}
and can be either a black-body spectrum at the current temperature, the stellar spectrum or uniform (in wavenumber).

A tolerance is set on the total error in a band, $\epsilon_\text{max}$. The number of $k$-terms in a band is chosen to be the smallest satisfying the criterion $\epsilon < \epsilon_\text{max}$. Once this division into subintervals has been made, the fitting of optimal $k$-coefficients is repeated for each $P$--$T$. These subsequent fits use the same subinterval spacing but with a reordering appropriate for each $P$--$T$.

In the tests presented here, we use two values for $\epsilon_\text{max}$: $\num{5e-3}$ and $\num{e-4}$, where $\epsilon_\text{max} = \num{e-4}$ is expected to reduce the error from the correlated-$k$ method significantly. We note, however, that the correlated-$k$ method will not completely converge to the LbL solution even with many $k$-coefficients (small $\epsilon_\text{max}$): $g(k)$ is calculated at each $P$--$T$ independently. If the absorption coefficient decreases with height at one wavelength and increases with height at a different wavelength within the same band, the two wavelengths will no longer correspond to the same value of $g$. A pseudo-monochromatic calculation where $g$ is kept constant is therefore not equivalent to a proper monochromatic calculation except in a very few special cases, see e.g. \citet{Goody1989} for more details.

The maximum column density over which $k$-coefficients are to be fitted must be determined. The column density of species $i$ is given by
\begin{equation}
u^i = \int_{z}^{\infty} \md z' \, \zeta_i (z') \rho(z')
= \frac{1}{g} \int_0^{P_{\tau=1}} \md P' \, \zeta_i (P')
= \frac{\zeta_i P}{g} ,
\label{eq:u^i}
\end{equation}
where we have assumed a constant mass mixing ratio. We set the maximum column density using \cref{eq:u^i} with the maximum pressure and mixing ratio in our $P$--$T$ table. In the ES radiation scheme, the mass mixing ratio used in \cref{eq:k_avg_tau=1} is calculated from the maximum column density using \cref{eq:u^i}.

To obtain the total flux in band $b$, $n_k$ pseudo-monochromatic calculations are performed, each yielding a total flux $F_l$. The band-integrated flux is then given by
\begin{equation}
F_b = \sum_{l=1}^{n_k} w_l^* F_l .
\end{equation}
The weight $w_l^*$ for the thermal component should ideally be identical to $w_l$ evaluated at the local temperature. For simplicity, however, the weights $w_l$ at the temperature where $\tau = 1$ are adopted. For the stellar component, $w_l^*$ is the stellar spectrum at the top of the atmosphere. We later compare these weighting schemes to a uniform weighting scheme, and show that, using the band structure adopted here, the exact weighting scheme does not affect fluxes and heating rates to a significant degree.

We use a band structure very similar to that used by \citet{Showman2009}, and we list our bands in \cref{tbl:bands}. Three small modifications have been made compared to the bands listed in \citet{Showman2009}: (i) The upper limit of band $30$ has been reduced from $\SI{38314}{\centi \metre^{-1}}$ to $\SI{28000}{\centi \metre^{-1}}$ to reduce the memory usage of our correlated-$k$ code, (ii) Band $31$ has been added to capture absorption up to the small wavelength limit of our line lists, and (iii) Band $32$ has been added to capture most of the stellar flux at small wavelengths.

\begin{table}
\centering
\caption{Table listing the bands used. They are almost identical to the bands in \citep{Showman2009}, the differences are explained in the text.}
\label{tbl:bands}
\begin{tabular}{r|r|r}
Band & Lower limit [$\si{\centi \metre^{-1}}$] & Upper limit [$\si{\centi \metre^{-1}}$] \\ \hline
$1$ & $\num{31}$ & $\num{217}$ \\
$2$ & $\num{217}$ & $\num{500}$ \\
$3$ & $\num{500}$ & $\num{962}$ \\
$4$ & $\num{962}$ & $\num{1550}$ \\
$5$ & $\num{1550}$ & $\num{1916}$ \\
$6$ & $\num{1916}$ & $\num{2273}$ \\
$7$ & $\num{2273}$ & $\num{2632}$ \\
$8$ & $\num{2632}$ & $\num{3041}$ \\
$9$ & $\num{3041}$ & $\num{3346}$ \\
$10$ & $\num{3346}$ & $\num{3992}$ \\
$11$ & $\num{3992}$ & $\num{4608}$ \\
$12$ & $\num{4608}$ & $\num{4950}$ \\
$13$ & $\num{4950}$ & $\num{5627}$ \\
$14$ & $\num{5627}$ & $\num{6277}$ \\
$15$ & $\num{6277}$ & $\num{6680}$ \\
$16$ & $\num{6680}$ & $\num{7519}$ \\
$17$ & $\num{7519}$ & $\num{8354}$ \\
$18$ & $\num{8354}$ & $\num{9091}$ \\
$19$ & $\num{9091}$ & $\num{9950}$ \\
$20$ & $\num{9950}$ & $\num{10417}$ \\
$21$ & $\num{10417}$ & $\num{10989}$ \\
$22$ & $\num{10989}$ & $\num{11628}$ \\
$23$ & $\num{11628}$ & $\num{12739}$ \\
$24$ & $\num{12739}$ & $\num{13423}$ \\
$25$ & $\num{13423}$ & $\num{14815}$ \\
$26$ & $\num{14815}$ & $\num{16340}$ \\
$27$ & $\num{16340}$ & $\num{17483}$ \\
$28$ & $\num{17483}$ & $\num{20202}$ \\
$29$ & $\num{20202}$ & $\num{25000}$ \\
$30$ & $\num{25000}$ & $\num{28000}$ \\
$31$ & $\num{28000}$ & $\num{31761}$ \\
$32$ & $\num{31761}$ & $\num{50000}$
\end{tabular}
\end{table}

\subsection{Band-averaged absorption coefficients}

Band-averaged absorption coefficients have been applied to exoplanet GCMs \citep{Dobbs-Dixon2013} and stellar/substellar atmosphere radiation hydrodynamical models \citep[see e.g][]{Freytag2010}. In \citet{Dobbs-Dixon2013}, an average absorption coefficient is calculated in each band as:
\begin{equation}
\bar k_b = \frac{\int_{\nut_b}^{\nut_{b+1}} \md \nut \, w(\nut) k_\rho(\nut)}{\int_{\nut_b}^{\nut_{b+1}} \md \nut \, w(\nut)},
\label{eq:k_b}
\end{equation}
where $\nut_b$ is the lower bound of bin $b$. The upper bound of the last band is defined as $\nut_{n_\text{b}+1}$, where $n_\text{b}$ is the number of bands. The fluxes $F_b$ are obtained by performing $n_\text{b}$ pseudo-monochromatic calculations, the total flux being the sum of the individual fluxes in each band. In \citet{Dobbs-Dixon2013}, the weighting function $w(\nut)$ is a black-body spectrum evaluated at the local temperature for the thermal component, i.e. $\bar k_b$ is the Planck mean in each band, known to be applicable in the optically thin limit. The stellar spectrum at the top of the atmosphere is used as weights for the stellar component. The bands were selected as in \citet{Showman2009}, see \cref{tbl:bands}.

Improved schemes utilising mean absorption coefficients exist, but we limit our discussion to the approach used by \citet{Dobbs-Dixon2013} and compare its accuracy to a full correlated-$k$ treatment. For this purpose, we show in \cref{sec:testing} results obtained with band-averaged absorption coefficients designed to replicate the treatment used in \citet{Dobbs-Dixon2013}.

\subsection{Atmo}

In order to investigate the accuracy of both the correlated-$k$ method and various two-stream approximations, we compare the ES radiation scheme to our line-by-line discrete ordinate code Atmo. It follows the method of the MARCS code \citep[for a description, see][]{Gustafsson2008}, with some modifications.

The 1D plane-parallel radiative transfer equation is solved using the discrete ordinates method \citep[see e.g.][]{Thomas2002}, i.e. solving the radiative transfer equation for discrete ray directions $\mu_i$, which are selected according to Gauss-Legendre quadrature. In this paper we use $16$ rays, and we have checked the convergence by using up to $32$ rays. The discrete ordinate equations are solved iteratively using the integral method, and the code has been parallelised to facilitate a high wavenumber resolution. By successively increasing the resolution, we found that a resolution on the order of $\sim \SI{e-3}{\centi \metre^{-1}}$ was necessary for the solution to have converged, i.e. about $\num{5e7}$ wavenumber points. In contrast, we use about $300$ pseudo-monochromatic calculations in the ES scheme, illustrating the large gain in computational efficiency achieved by using the correlated-$k$ method.

\section{Testing the correlated-\texorpdfstring{$k$}{k} and two-stream approximations} \label{sec:testing}

In this section we test the accuracy of the ES radiation scheme by comparing it to Atmo. The accuracy is investigated for several different scenarios designed to test a range of physical conditions representative for hot Jupiters.

To ease comparison between the different two-stream approximations, we list the $L^1$ norm of the error,
\begin{align}
L^1 &= \int_{\log_{10} P_\text{min}}^{\log_{10} P_\text{max}} (\md \log_{10} P) \left[ \frac{\left|\mathcal H_\text{Atmo}\right|}{\int_{\log_{10} P_\text{min}}^{\log_{10} P_\text{max}} (\md \log_{10} P) \, \left|\mathcal H_\text{Atmo}\right|} \right. \notag \\
&\quad \left. \times
\frac{\left|\mathcal H_\text{ES} - \mathcal H_\text{Atmo}\right|}{\left|\mathcal H_\text{Atmo}\right|} \right] \notag \\
&= \frac{\int_{\log_{10} P_\text{min}}^{\log_{10} P_\text{max}} (\md \log_{10} P) \, \left|\mathcal H_\text{ES} - \mathcal H_\text{Atmo}\right|}{\int_{\log_{10} P_\text{min}}^{\log_{10} P_\text{max}} (\md \log_{10} P) \, \left|\mathcal H_\text{Atmo}\right|},
\end{align}
given here for the heating rate, where $\mathcal H_\text{ES}$ and $\mathcal H_\text{Atmo}$ are the heating rates from the ES radiation scheme and Atmo, respectively, and $P_\text{min}$ ($P_\text{max}$) is the minimum (maximum) pressure in our calculations. This is a convenient measure to use when comparing errors between different two-stream approximations and opacity treatments, and represents the relative error of some quantity, in this case the heating rate, weighted by the current value of that quantity integrated over all pressures.

In \cref{ap:test0}, we describe a very simple test where analytical solutions to both the two-stream approximated and the full radiative transfer equation exist. A grey opacity is used to eliminate errors from the correlated-$k$ method. This scenario is used to test both the accuracy of the numerical solvers and the two-stream approximation in isolation. The test confirms that the numerical solvers of both the ES radiation scheme and Atmo have satisfactory accuracies. The value of $D$ giving the most accurate results ($D = 1.66$) yields an error of about $\SI{1}{\percent}$ for the flux and $\SI{10}{\percent}$ for the heating rate.

We have also performed a test designed to be more realistic, but still minimising the error caused by the correlated-$k$ method by including only H$_2$--H$_2$ collision induced (continuum) absorption (CIA) in an isothermal atmosphere without irradiation. For brevity, we do not describe the test here, but only summarise the main results. We found that the value of $D$ giving the most accurate results, $D = 1.66$, yielded an error in the flux of $<\SI{1}{\percent}$, and about $\SI{7.5}{\percent}$ in the heating rate using $\epsilon = \num{5e-3}$ ($1$ to $2$ $k$-coefficients in each band). Using a mean absorption coefficient in each band yields similar errors.

In the next sections, we describe three other tests in detail. Test 1 includes only absorption by H$_2$O, a very important absorber in the atmospheres of both the Earth and hot Jupiters. Tests 2 and 3 are designed to test the accuracy for a typical well-mixed hot Jupiter night-side and day-side, respectively, where the day-side includes irradiation and absorption by TiO and VO in the upper atmosphere. A gravitational acceleration of $\SI{9.42}{\metre \per \second^2}$ is used, suitable for HD~209458b. Unless otherwise stated, $k$-coefficients were calculated using $\epsilon_\text{max} = \num{5e-3}$, a Planckian weighting scheme for the thermal component and the stellar spectrum for the stellar component, unless stated otherwise.

\subsection{Test 1: Pure H\texorpdfstring{$_2$}{2}O absorption in a high-temperature isothermal atmosphere} \label{sec:test1}

This test includes only absorption by H$_2$O. The temperature is fixed to $\SI{1500}{\kelvin}$ and the atmospheric domain extends from $\SI{e-1}{\pascal}$ to $\SI{e8}{\pascal}$, using $100$ pressure points on a logarithmic scale. Irradiation at the upper boundary is not included and the lower boundary emits as a black body at $\SI{1500}{\kelvin}$ with zero albedo. A constant mass mixing ratio of $3.3477\e{-3}$ is adopted, which corresponds to the smallest mass mixing ratio predicted by Eq.~(A4) in \citep{Burrows1999}. Adopting $\epsilon_\text{max}=\num{5e-3}$ and $\epsilon_\text{max}=\num{e-4}$ yield $\sim 10$ and $\sim 100$ $k$-coefficients in each band, respectively. Note, however, that the number of $k$-coefficients can vary significantly between different bands.

\begin{figure}
\includegraphics[width=\columnwidth]{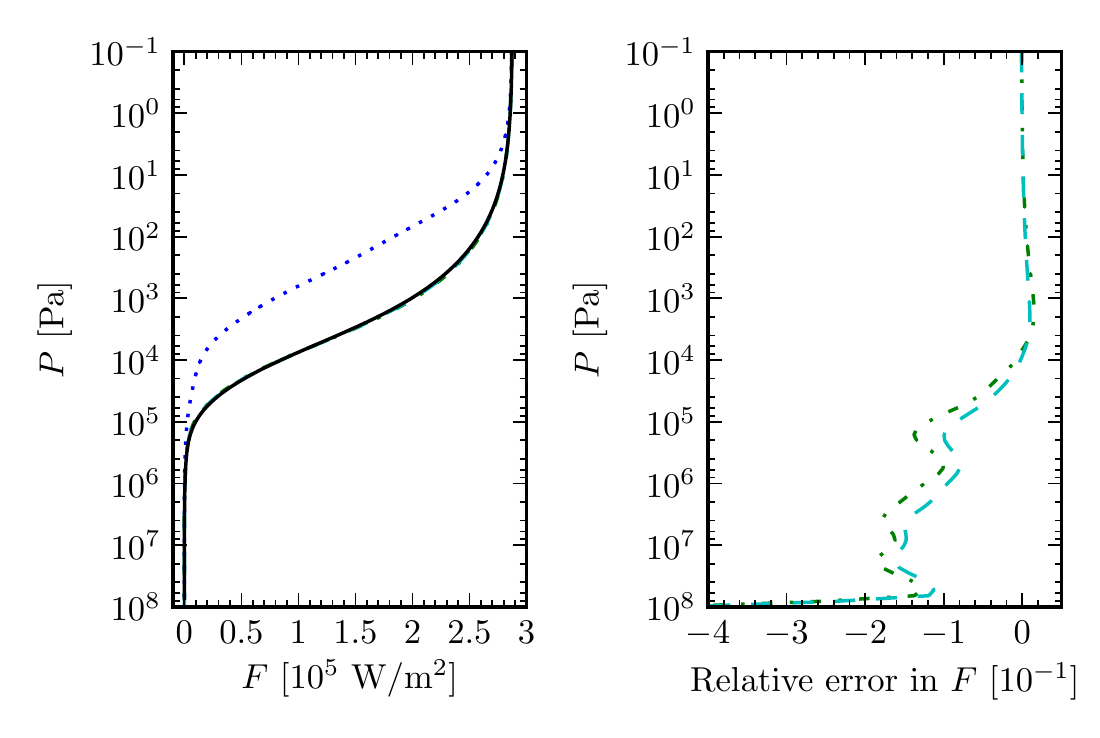}
\caption{The left-hand panel shows the fluxes obtained with ES radiation scheme using the two-stream approximation and correlated-$k$ method obtained with $D = 1.66$ and $\epsilon_\text{max} = \num{5e-3}$ (dashed-dotted, green), $\epsilon_\text{max} = \num{e-4}$ (dashed, cyan), and mean absorption coefficients (dotted, blue), for an isothermal atmosphere with pure H$_2$O absorption. The Atmo LbL DO result is also shown in this panel (solid, black) and is used to calculate the relative errors shown in the right-hand panel (except for the mean absorption coefficient case since errors are too large). Relative errors stay below $\SI{30}{\percent}$ throughout the atmosphere.}
\label{fig:test1_nflx}
\end{figure}

\begin{figure}
\includegraphics[width=\columnwidth]{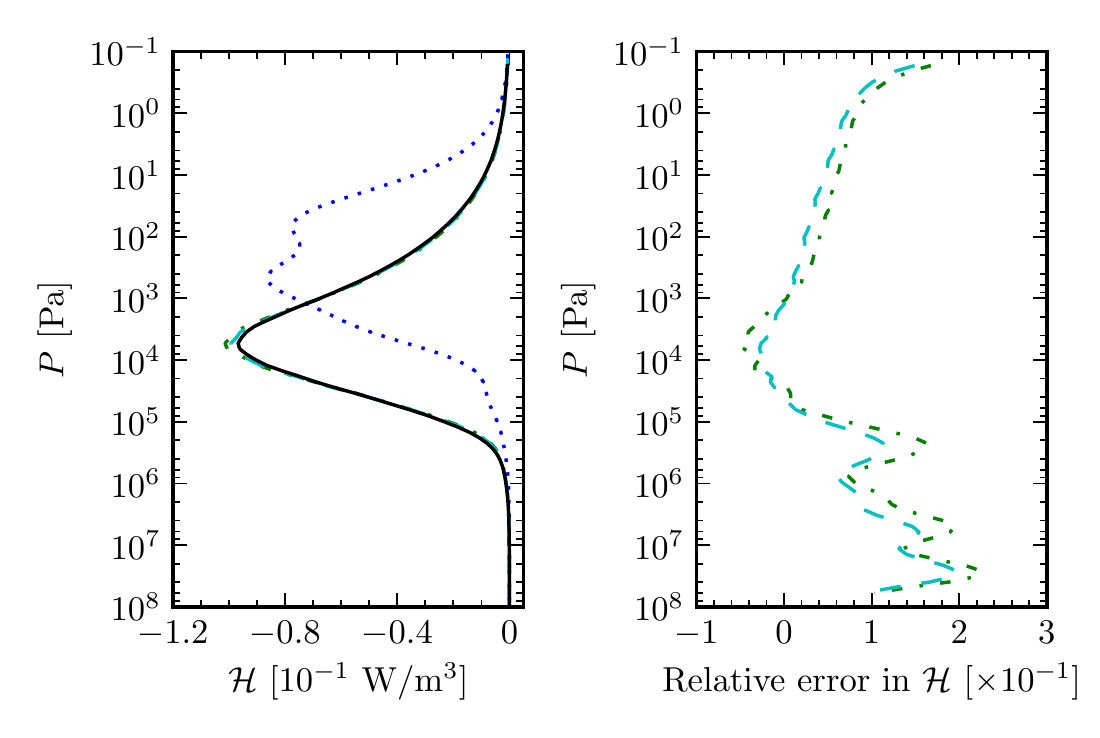}
\caption{Same as \cref{fig:test1_nflx}, but for heating rates. Relative errors stay below $\SI{30}{\percent}$ throughout the atmosphere. Note that in the region where the heating rate (magnitude) is large, the error remains small.}
\label{fig:test1_hrts}
\end{figure}

\Cref{fig:test1_nflx,fig:test1_hrts} show the fluxes and heating rates, respectively, with corresponding relative errors. Since the atmosphere is isothermal, the upward flux is the Planck flux throughout the atmosphere. At the upper boundary, the downward flux is $0$, while at the lower boundary it is the Planck flux due to the high optical depth. The total flux is therefore $0$ at high optical depths, while at low optical depths the planet radiates as a black body, as expected, with a heating rate peak at $\sim \SI{e4}{\pascal}$. Errors in the heating rate generally stay below about $\SI{10}{\percent}$ at pressures where the heating is significant, while using $\epsilon_\text{max} = \num{e-4}$ yields significantly more accurate results. \cref{tbl:test1} shows $L^1$ errors, indicating that $D=1.66$ and $\sqrt{3}$ yield the most accurate results. There is no significant difference between a Planckian and a uniform weighting (UW) scheme.

Using mean absorption coefficients (see \cref{fig:test1_nflx,fig:test1_hrts,tbl:test1}) yield very inaccurate fluxes and heating rates. The flux is underestimated at a given pressure, while the heating rate peak occurs at pressures about two orders of magnitude smaller than the peak of the LbL DO result. The fact that one $k$-coefficient per band is not sufficient to resolve the opacity is reflected by the need for $\sim 10$ $k$-coefficients in each band to achieve a tolerance of $\epsilon_\text{max}=\num{5e-3}$ for H$_2$O. The failure of this method is discussed in more detail in \cref{sec:DD}.

\begin{table}
\centering
\caption{Computed flux ($F$) and heating rate ($\mathcal H$) $L^1$ norms for test~1, showing that the most accurate fluxes and heating rates are obtained with $D = 1.66$ and $\sqrt{3}$. The last two rows correspond to a uniform weighting scheme (UW) and band-averaged (mean) absorption coefficients, respectively.}
\begin{tabular}{l|r|r}
 & $L^1$, $F$ & $L^1$, $\mathcal H$ \\ \hline
$D = \sqrt{3}$ & $0.007$ & $0.044$ \\
$D = 1.66$ & $0.007$ & $0.046$ \\
$D = 2$ & $0.013$ & $0.063$ \\
$D = 1.66$, $\epsilon_\mathrm{max}=10^{-4}$ & $0.004$ & $0.021$ \\
$D = 1.66$, UW & $0.007$ & $0.044$ \\
$D=1.66$, mean & $0.227$ & $0.837$
\end{tabular}
\label{tbl:test1}
\end{table}

\subsection{Test 2: Mixed night-side hot Jupiter atmosphere}

\begin{figure}
\includegraphics[width=\columnwidth]{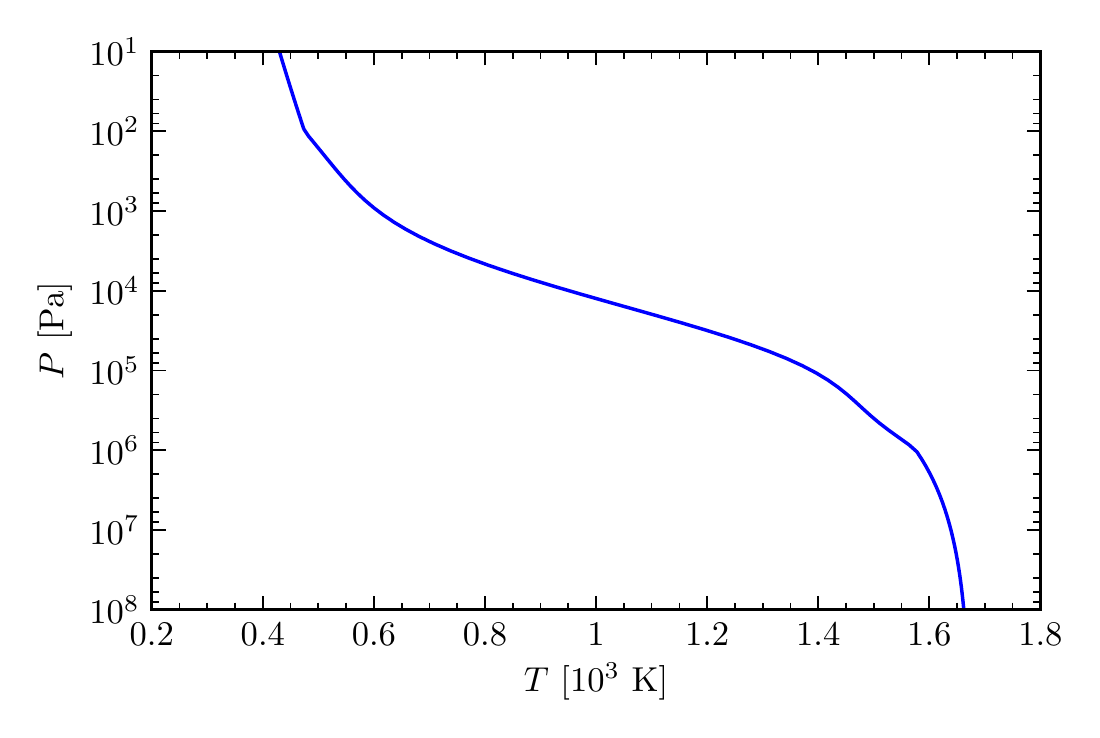}
\caption{The $P$--$T$ profile used in test~2. From the polynomial fit \citep{Heng2011} to the night-side profile of HD~209458b from \citet{Iro2005} with the smoothing described in \citet{Mayne2013b}. The temperature varies from about $\SI{400}{\kelvin}$ high in the atmosphere to above $\SI{1600}{\kelvin}$ in the deeper layers.}
\label{fig:test2_pt}
\end{figure}

We next consider conditions representative of a real hot Jupiter atmosphere. We use a polynomial fit \citep{Heng2011} to the night-side $P$--$T$ profile from \citet{Iro2005} with the smoothing described in \citet{Mayne2013b}, shown in \cref{fig:test2_pt}. The temperature varies from about $\SI{400}{\kelvin}$ in the upper atmosphere to above $\SI{1600}{\kelvin}$ at $\SI{e8}{\pascal}$, consistent with the literature (see e.g. Fig.~6 in \citet{Showman2009} and Fig.~7 in \citet{Baraffe2008}). Irradiation at the upper boundary is not included and the lower boundary emits as a black body with $T_\text{lb} = T(P_\text{lb})$, where $T_\text{lb}$ and $P_\text{lb}$ are the temperature and pressure at the lower boundary, respectively. From the $P$--$T$ profile in \cref{fig:test2_pt}, $T_\text{lb} = \SI{1662}{\kelvin}$. All molecules in \cref{tbl:opacity_data} are included except TiO and VO, with abundances as described in \cref{sec:abundances}.

\begin{figure}
\includegraphics[width=\columnwidth]{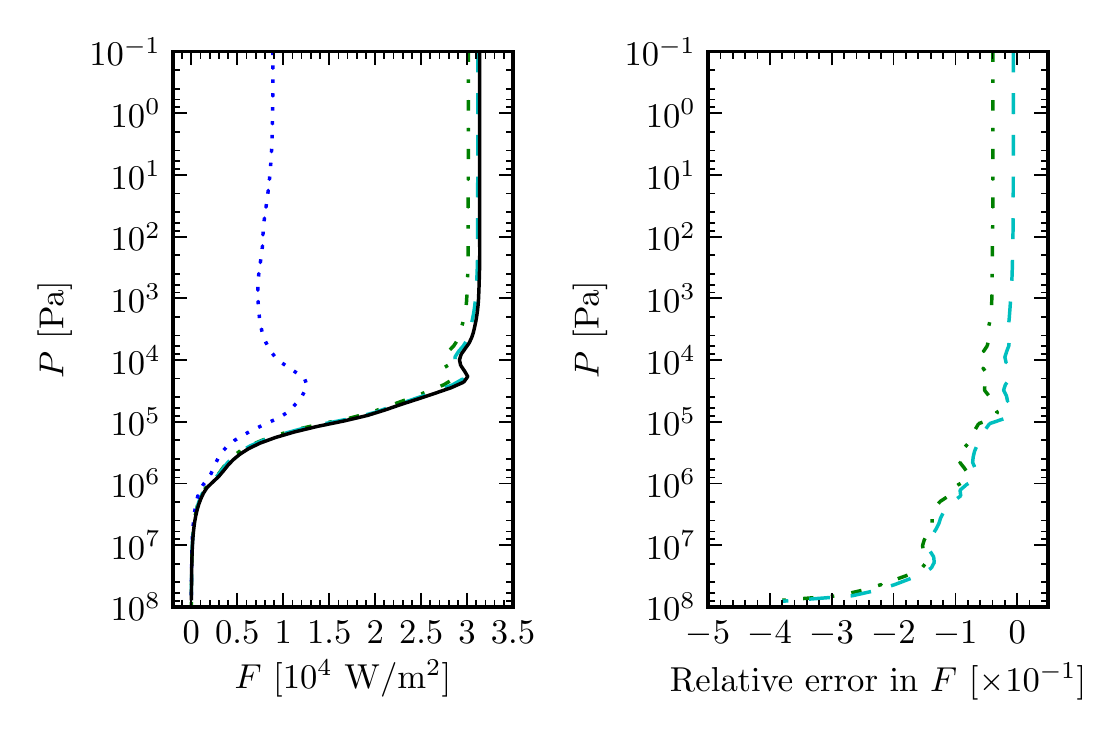}
\caption{The left-hand panel shows the fluxes obtained in test~2 with the ES radiation scheme using the $P$--$T$ profile in \cref{fig:test2_pt}. All opacity sources listed in \cref{tbl:opacity_data} are included except TiO and VO. The Atmo LbL DO result is also shown in this panel and is used to calculate the relative errors shown in the right-hand panel.
For explanation of the different lines, see the caption of \cref{fig:test1_nflx}.}
\label{fig:test2_nflx}
\end{figure}

\begin{figure}
\includegraphics[width=\columnwidth]{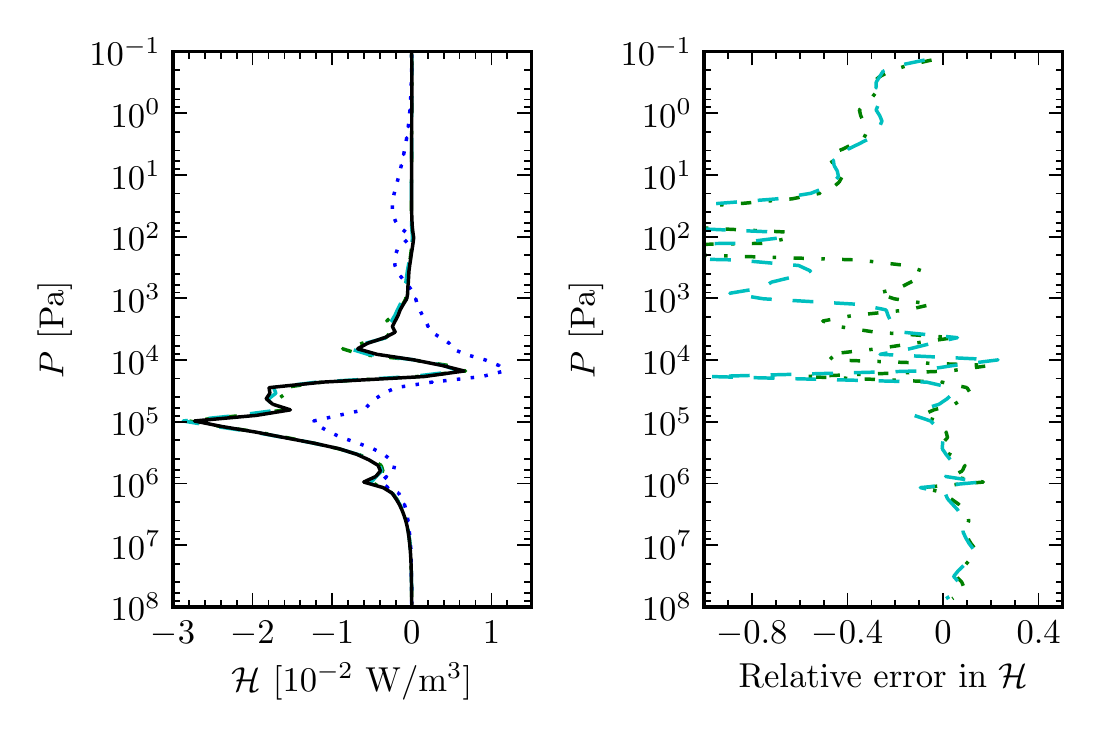}
\caption{Same as \cref{fig:test2_nflx} for the heating rates. Note that when the heating rate is very small relative errors may become large. The effect on the heating budget will be small, however, so we do not consider this a problem.
}
\label{fig:test2_hrts}
\end{figure}

Fluxes and heating rates with relative errors are plotted in \cref{fig:test2_nflx,fig:test2_hrts}, with $L^1$ errors given in \cref{tbl:test2}, from which it is clear that $D = 1.66$ yields the most accurate fluxes and heating rates overall.

\begin{table}
\centering
\caption{Computed flux ($F$) and heating rate ($\mathcal H$) $L^1$ norms for test~2, showing that the most accurate fluxes and heating rates are obtained with $D = 1.66$.}
\begin{tabular}{l|r|r}
 & $L^1$, $F$ & $L^1$, $\mathcal H$ \\ \hline
$D = \sqrt{3}$ & $0.065$ & $0.096$ \\
$D = 1.66$ & $0.043$ & $0.080$ \\
$D = 2$ & $0.133$ & $0.173$ \\
$D = 1.66$, $\epsilon_\mathrm{max}=10^{-4}$ & $0.012$ & $0.070$ \\
$D = 1.66$, UW & $0.031$ & $0.081$ \\
$D=1.66$, mean & $0.700$ & $0.926$
\end{tabular}
\label{tbl:test2}
\end{table}

When the flux or heating rate is close to zero, the relative error can become large. We do not consider this as a problem, however, as it does not have a large impact on the atmospheric heat budget. The overall results are similar to those obtained in \cref{sec:test1}. Note that calculations with mean absorption coefficients in each band significantly underestimates the flux and result in heating rate peaks with the wrong amplitude.

\subsection{Test 3: Mixed day-side hot Jupiter atmosphere}

\begin{figure}
\includegraphics[width=\columnwidth]{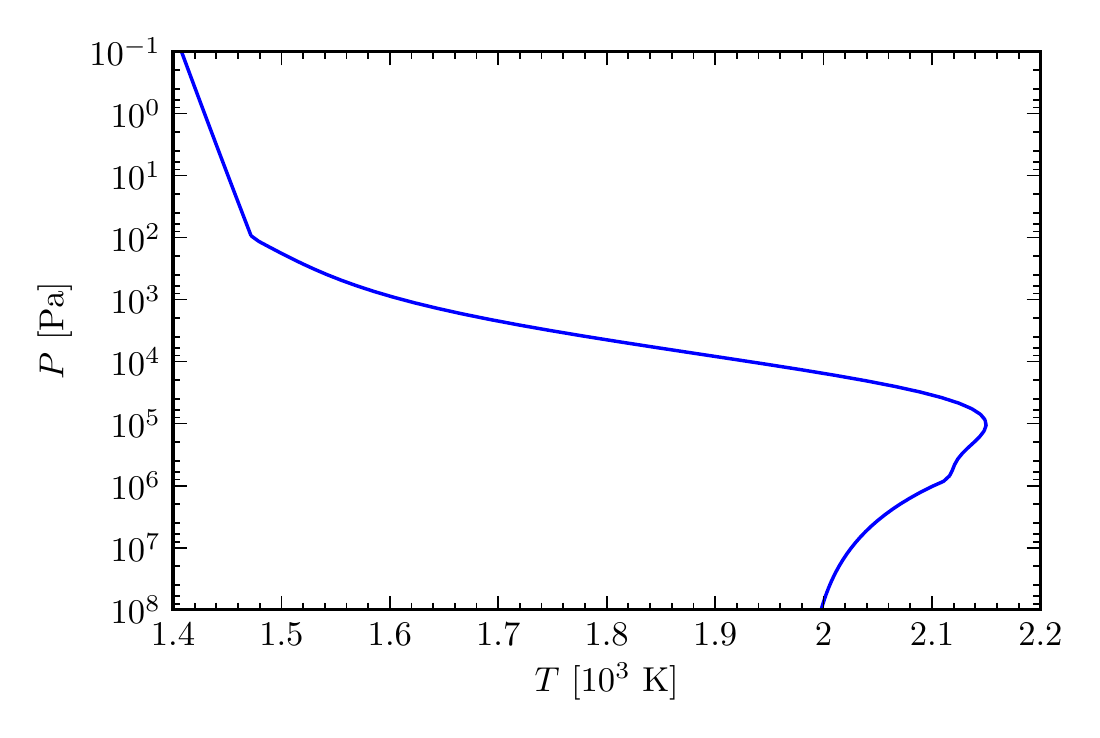}
\caption{The $P$--$T$ profile used in test~3. From the polynomial fit \citep{Heng2011} to the day-side profile of HD~209458b from \citet{Iro2005} with the smoothing described in \citet{Mayne2013b}. The temperature varies from about $\SI{1400}{\kelvin}$ in the upper atmosphere to about $\SI{2000}{\kelvin}$ in the deeper layers.}
\label{fig:test3_pt}
\end{figure}

Our last test adopts conditions suitable to the day-side of hot Jupiters. We use the polynomial fit \citep{Heng2011} to the day-side $P$--$T$ profile of HD 209458b from \citet{Iro2005} with the smoothing described in \citet{Mayne2013b}, plotted in \cref{fig:test3_pt}, consistent with the literature~\citep{Showman2009,Baraffe2008}. The thermal and stellar components of the flux are calculated separately and then summed to obtain the total flux and heating rate. For the thermal component, the lower boundary again emits as a black body at $T_\text{lb} = T(P_\text{lb})$, i.e. $\SI{1998}{\kelvin}$ using the $P$--$T$ profile in \cref{fig:test3_pt}. Note that, due to the separation of the intensity into direct and diffuse components and the our neglect of scattering, the stellar component of the intensity will only be subject to errors caused by the correlated-$k$ method.

We assume an orbital distance $a_\text{orbit} = \SI{0.047}{\astronomicalunit}$ and a parent star effective temperature $T_\text{eff}^\text{star} = T_\text{eff}^\text{Sun} = \SI{5785}{\kelvin}$ and radius $R_\text{star} = R_\text{Sun}$. Using Stefan-Boltzmann's law, the stellar irradiation at the top of the planet's atmosphere is given by
\begin{equation}
F_\text{TOA}^\text{star} = \sigma \left(T_\text{eff}^\text{star} \right)^4 \left( \frac{R_\text{star}}{a_\text{orbit}} \right)^2 = \SI{6.092e5}{\watt \per \metre^2} .
\end{equation}
We adopt a zero solar zenith angle and use a solar spectrum from Kurucz\footnote{See \url{http://kurucz.harvard.edu/}.}. At smaller wavelengths than available, we set the stellar flux to zero, while at larger wavelengths we extrapolate using a black-body spectrum with the effective temperature of the Sun ($T = \SI{5785}{\kelvin}$).

\begin{figure}[tp]
\includegraphics[width=\columnwidth]{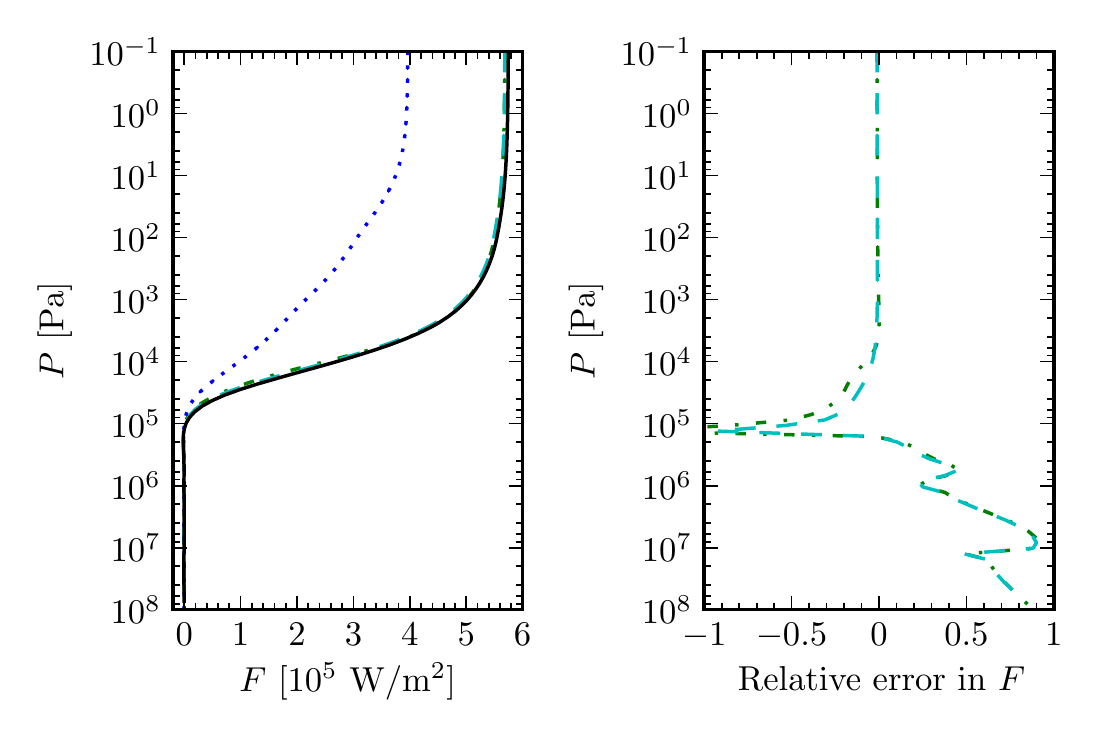}
\caption{The left-hand panel shows the \emph{thermal component} of the flux as a function of total pressure for test~3. The Atmo LbL DO result is also shown in this panel and is used to calculate the relative errors shown in the right-hand panel. For explanation of the different lines, see the caption of \cref{fig:test1_nflx}. At high pressures the relative error becomes large, similar to that seen in \cref{fig:test2_hrts} for the heating rate. As the flux itself is small, however, this is not a problem.}
\label{fig:test3_nflx_lw}
\end{figure}

\begin{figure}[tp]
\includegraphics[width=\columnwidth]{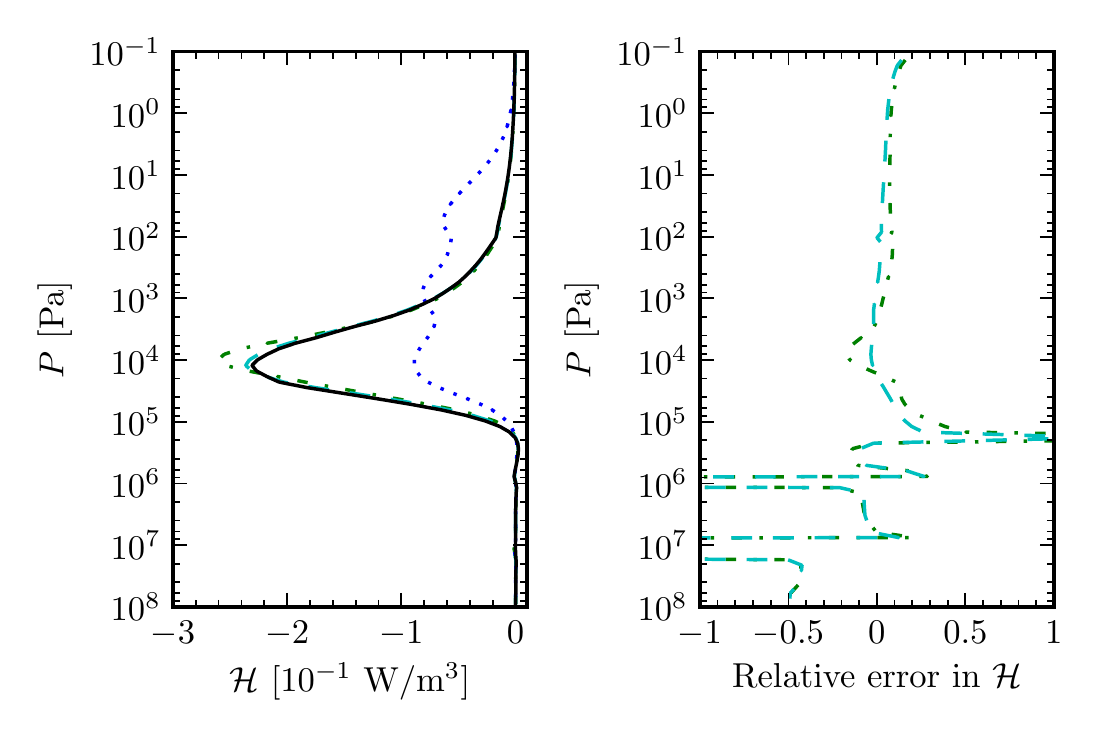}
\caption{Same as \cref{fig:test3_nflx_lw} for the \emph{thermal component} of the heating rate. Relative errors become unreasonably large only where the heating rate is very small.}
\label{fig:test3_hrts_lw}
\end{figure}

\Cref{fig:test3_nflx_lw,fig:test3_hrts_lw} show the \emph{thermal} flux and heating rate with relative errors. The flux error is small in regions with non-negligible flux. The heating rate error also remains small in regions with significant cooling. This is confirmed by the computed $L^1$ norms listed in \cref{tbl:test3_lw}. A diffusitivy of $D = 1.66$ yields the smallest error, and it is approximately halved by decreasing $\epsilon_\text{max}$ to $\num{e-4}$. Whether a black-body or uniform weighting scheme is used does not affect the accuracy significantly, but using a mean absorption coefficient does result in significant errors.

\begin{table}
\centering
\caption{Computed flux ($F$) and heating rate ($\mathcal H$) $L^1$ norms for the \emph{thermal component} in test~3. The smallest errors are again obtained with $D = 1.66$.}
\begin{tabular}{l|r|r}
 & $L^1$, $F$ & $L^1$, $\mathcal H$ \\ \hline
$D = \sqrt{3}$ & $0.026$ & $0.100$ \\
$D = 1.66$ & $0.015$ & $0.097$ \\
$D = 2$ & $0.064$ & $0.128$ \\
$D = 1.66$, $\epsilon_\mathrm{max}=10^{-4}$ & $0.014$ & $0.033$ \\
$D = 1.66$, UW & $0.012$ & $0.097$ \\
$D=1.66$, mean & $0.425$ & $0.711$
\end{tabular}
\label{tbl:test3_lw}
\end{table}

\begin{figure}[tp]
\includegraphics[width=\columnwidth]{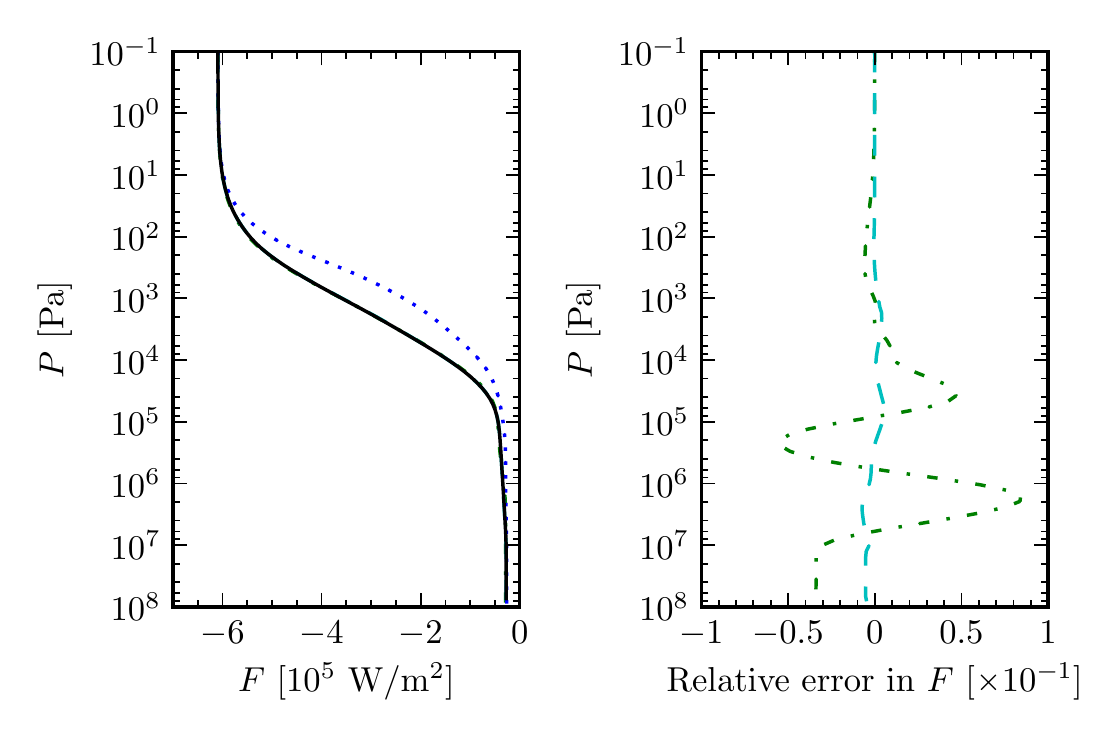}
\caption{The left-hand panel shows the \emph{stellar component} of the flux as a function of total pressure for test~3 obtained with $\epsilon_\mathrm{max} = 5 \times 10^{-3}$ (dashed-dotted, green) and $\epsilon_\mathrm{max} = 10^{-4}$ (dashed, cyan). The Atmo LbL DO result is also shown in this panel (solid, black) and is used to calculate the relative errors shown in the right-hand panel. Errors are small, and using $\epsilon_\text{max} = \num{e-4}$ almost completely eliminates errors in the ES radiation scheme.}
\label{fig:test3_nflx_sw}
\end{figure}

\begin{figure}[tp]
\includegraphics[width=\columnwidth]{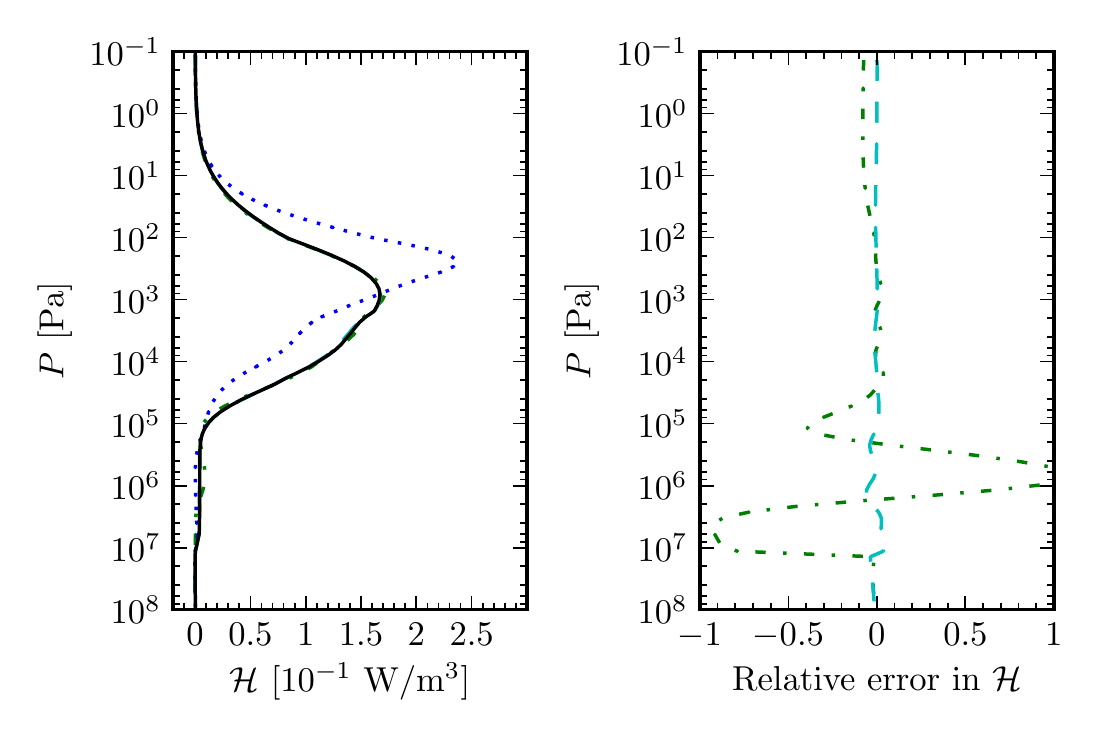}
\caption{Same as \cref{fig:test3_nflx_sw} for the \emph{stellar component} of the heating rate.}
\label{fig:test3_hrts_sw}
\end{figure}

\Cref{fig:test3_nflx_sw,fig:test3_hrts_sw} show the \emph{stellar} flux and heating rate with relative errors. At the top of the atmosphere, the flux is $-\SI{6.092e5}{\watt \per \metre^2}$, as prescribed, and is subsequently absorbed. The heating rate is positive, i.e. the atmosphere is heated due to the absorption of stellar radiation, as expected. The accuracy is acceptable, the error in the flux stays below $\SI{10}{\percent}$, while the heating rate error also stays below $\SI{10}{\percent}$ in the regions with strong heating. This is reflected in the $L^1$ errors listed in \cref{tbl:test3_sw}. Using $\epsilon_\text{max} = \num{e-4}$ significantly reduces the error from the correlated-$k$ method, and changing the weighting scheme does not alter the results significantly. The use of an average absorption coefficient, however, is seen to still result in significant errors. See \cref{sec:DD} for discussion and more details.

\begin{table}
\centering
\caption{Computed flux ($F$) and heating rate ($\mathcal H$) $L^1$ norms for the \emph{stellar component} in test~3. The correlated-$k$ method is seen to introduce errors of about $\SI{4}{\percent}$.}
\begin{tabular}{l|r|r}
 & $L^1$, $F$ & $L^1$, $\mathcal H$ \\ \hline
$\epsilon_\mathrm{max} = 5 \times 10^{-3}$ & $0.004$ & $0.035$ \\
$\epsilon_\mathrm{max} = 10^{-4}$ & $0.001$ & $0.005$ \\
$\epsilon_\mathrm{max} = 5 \times 10^{-3}$, UW & $0.006$ & $0.045$ \\
Mean & $0.094$ & $0.432$
\end{tabular}
\label{tbl:test3_sw}
\end{table}

\begin{figure}[tp]
\includegraphics[width=\columnwidth]{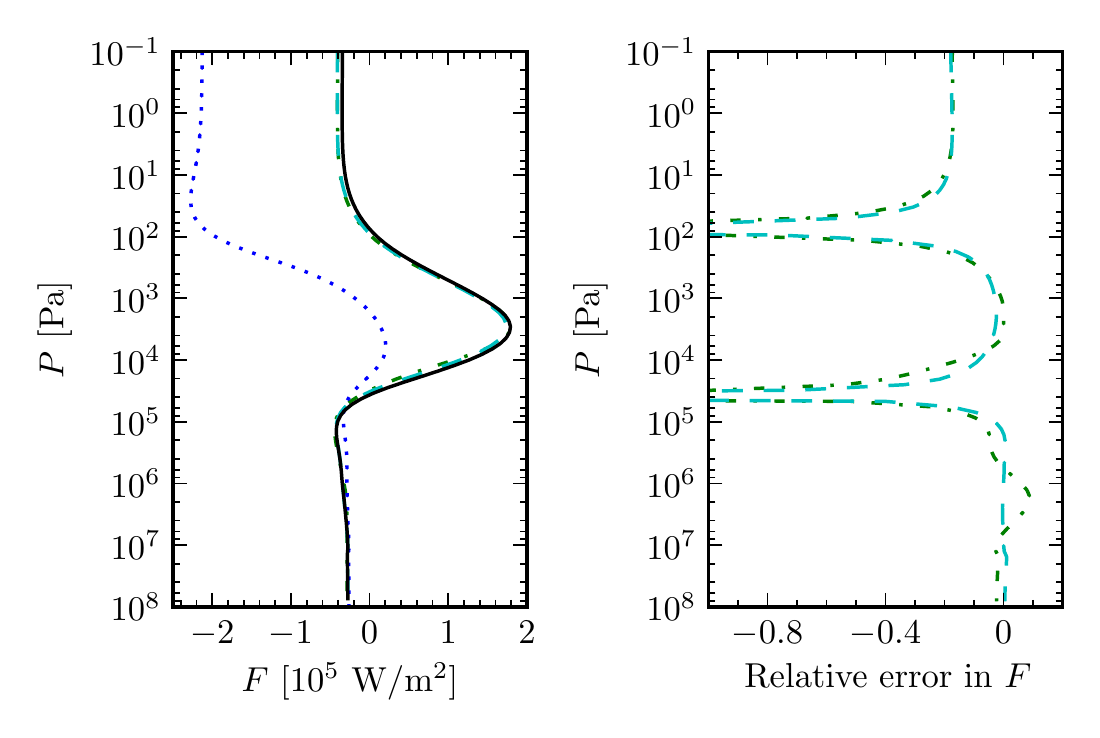}
\caption{The left-hand panel shows the \emph{total} flux as a function of total pressure for test~3. The Atmo LbL DO result is also shown in this panel and is used to calculate the relative errors shown in the right-hand panel. For explanation of the different lines, see the caption of \cref{fig:test1_nflx}. Again, relative errors become unreasonably large only where the flux is very small.}
\label{fig:test3_nflx}
\end{figure}

\begin{figure}[tp]
\includegraphics[width=\columnwidth]{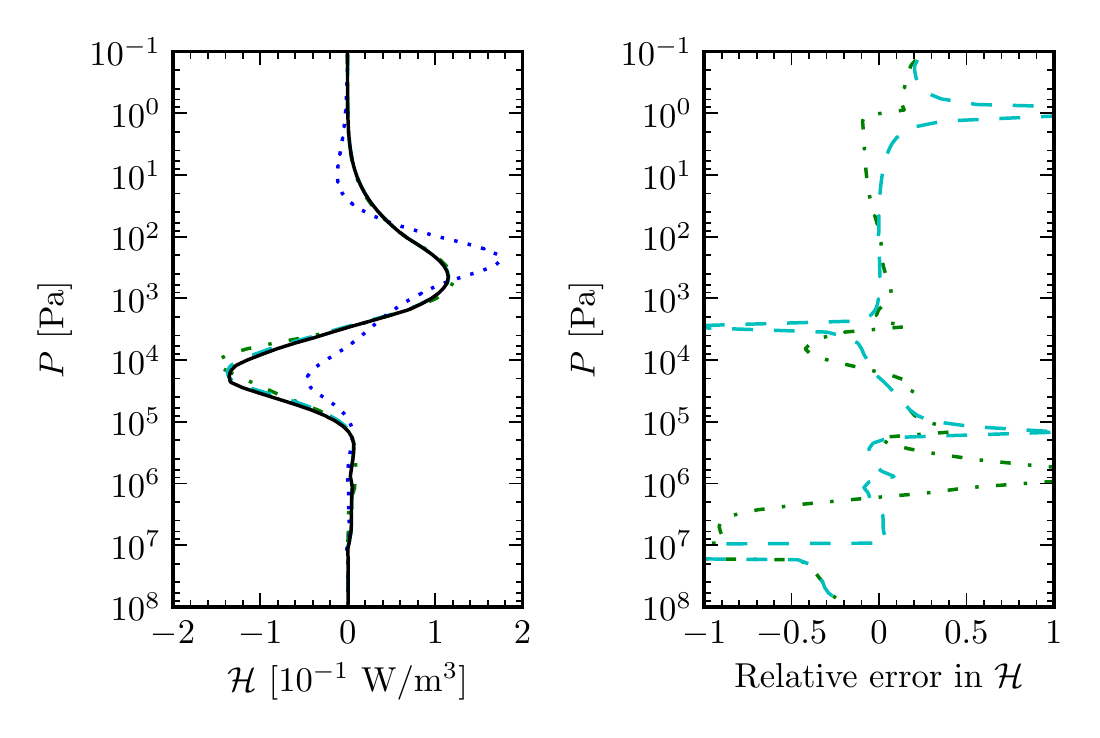}
\caption{Same as \cref{fig:test3_nflx} for the \emph{total} heating rate. Relative errors become unreasonably large only where the heating rate is very small, with a negligible effect on the heating budget.
}
\label{fig:test3_hrts}
\end{figure}

The \emph{total} flux and heating rate, obtained by summing up the stellar and thermal components of the flux and heating rate, are shown in \cref{fig:test3_nflx,fig:test3_hrts}, respectively. The main region of heating and cooling, seen separately in \cref{fig:test3_hrts_lw,fig:test3_hrts_sw}, respectively, are still clearly distinguishable in \cref{fig:test3_hrts}. The atmosphere is heated at low pressures and cooled slightly at higher pressures. Note that errors remain satisfactory small for relevant (i.e. non zero) values of the heating rate, as also shown in \cref{tbl:test3}. The error introduced by using mean absorption coefficients is significant.

\begin{table}
\centering
\caption{Computed flux ($F$) and heating rate ($\mathcal H$) $L^1$ norms for the \emph{total} flux and heating rate in test~3. Again $D = 1.66$ yields the smallest errors.}
\begin{tabular}{l|r|r}
 & $L^1$, $F$ & $L^1$, $\mathcal H$ \\ \hline
$D = \sqrt{3}$ & $0.164$ & $0.132$ \\
$D = 1.66$ & $0.097$ & $0.124$ \\
$D = 2$ & $0.387$ & $0.169$ \\
$D = 1.66$, $\epsilon_\mathrm{max}=10^{-4}$ & $0.081$ & $0.043$ \\
$D = 1.66$, UW & $0.090$ & $0.116$ \\
$D=1.66$, mean & $2.034$ & $0.624$
\end{tabular}
\label{tbl:test3}
\end{table}

\subsection{Discussion of the failure of mean absorption coefficients} \label{sec:DD}

Inspection of \cref{fig:test1_nflx,fig:test1_hrts,fig:test2_nflx,fig:test2_hrts,fig:test3_nflx_sw,%
fig:test3_hrts_sw,fig:test3_nflx_lw,fig:test3_hrts_lw,fig:test3_nflx,fig:test3_hrts} suggests systematic deviations of results based on the band-averaged absorption coefficients. For a given pressure, the thermal and stellar fluxes are underestimated, often resulting in heating rate peaks occurring at lower pressures and having the wrong magnitude. In an attempt to explain this behaviour, we first consider the direct stellar component.

The band-integrated direct stellar component of the flux is given by
\begin{equation}
F_{\text{s},b}^- (u_\rho) = \frac{1}{\mu_0} \int_{\nut_b}^{\nut_{b+1}} \md \nut \, F_\text{s} e^{-k_\rho(\nut) u_\rho/\mu_0},
\label{eq:F_s_b_LbL}
\end{equation}
where the atmospheric slab has been assumed to be homogeneous where $u_\rho$ is the mass column density down to some height $z$. Using a mean absorption coefficient instead, the corresponding flux is
\begin{equation}
F_{\text{s},b}^- (u_\rho) = \frac{e^{-\bar k_b u_\rho/\mu_0}}{\mu_0} \int_{\nut_b}^{\nut_{b+1}} \md \nut \, F_\text{s} .
\label{eq:F_s_b_mean}
\end{equation}
For simplicity we assume the incoming stellar radiation at the top of the atmosphere is wavenumber independent within a given band. Using a mean absorption coefficient then implies
\begin{equation}
\int_{\nut_b}^{\nut_{b+1}} \md \nut \, e^{-k_\rho(\nut) u_\rho/\mu_0} \approx (\nut_{b+1} - \nut_b) e^{-\bar k_b u_\rho/\mu_0} .
\end{equation}
Within a band the absorption coefficient $k_\rho(\nut)$ will vary by orders of magnitude, causing some regions in the band to have a small transmission and others to have a large transmission. The mean in \cref{eq:k_b} is an arithmetic mean, i.e. the largest values of $k_\rho(\nut)$ will dominate $\bar k_b$. Regions with high transmission due to small $k_\rho(\nut)$ will be overshadowed by this large mean, causing the overall transmission to be underestimated. This explains the deviation in the flux seen in \cref{fig:test3_nflx_sw}.

A similar argument can be used in the thermal region, but upward and downward radiation need to be considered separately. The radiative transfer equation reads
\begin{equation}
\frac{\md I_\nut}{\md s} = k(\nut,s) \left[ B_\nut(s) - I_\nut(s) \right],
\label{eq:RT_I(s)}
\end{equation}
where $s$ is the path over which radiation travels. We first consider the isothermal case, where the upward radiation is constant and equal to the black-body flux throughout the atmosphere. At the top of the atmosphere, the downward flux is zero, i.e. the change in intensity will, according to \cref{eq:RT_I(s)}, be dominated by thermal emission ($B_\nut(s) > I_\nut(s)$). Using a band-mean absorption coefficient effectively increases $k(\nut,s)$, which in \cref{eq:RT_I(s)} yields a larger intensity at a given $s$ or pressure. The downward radiation contributes negatively to the total flux, i.e. the total flux will be smaller for a given pressure, as seen in \cref{fig:test1_nflx}.

If the atmosphere has non-zero temperature gradients, the upward flux will also depend on pressure. In both $P$--$T$ profiles used here, the temperature decreases with height overall. At the lower boundary the upward intensity is simply the Planck intensity, i.e. the right-hand side of \cref{eq:RT_I(s)} is zero. As the temperature decreases, $B_\nut(s)$ will generally decrease, causing $B_\nut(s) < I_\nut(s)$. The upward flux is therefore dominated by absorption, and effectively increasing $k_\rho(\nut)$ will cause the upward flux to become smaller for a given $s$ or pressure. This explains why the total flux at the top of the atmosphere is underestimated when using a band-mean absorption coefficient, as seen in \cref{fig:test1_nflx,fig:test2_nflx,fig:test3_nflx}.

These results show that large errors in both fluxes and heating rates may occur when the mean opacity scheme described in \citet{Dobbs-Dixon2013} is applied in hot Jupiter GCMs. Improved mean opacity schemes have been developed by the stellar atmosphere community \citep[see e.g.][]{Nordlund1982,Skartlien2000}, which may be applicable to hot Jupiter atmospheres. Further developments of these improved schemes may be needed, however, as they rely on correlations induced by strong vertical stratification, and longitude--latitude-dependent stellar heating has not been considered. Further discussion is, however, beyond the scope of the present work.

\section{Conclusions} \label{sec:conclusions}

The accuracy of radiation schemes used in GCMs has been studied extensively for Earth-like conditions, but detailed analysis for hot Jupiter-like conditions are lacking. In this paper we have analysed the accuracy and uncertainties in state-of-the-art radiation schemes used in several GCMs applied to hot Jupiters. Opacity sources and calculation of absorption coefficients from high-temperature line lists have been discussed. We present a line profile cut-off scheme that decreases the computation time required to calculate absorption coefficients by a factor of $\sim 100$ compared to other methods used in the literature, while still giving accurate results. Both the two-stream approximation and correlated-$k$ method's applicability to hot Jupiter atmospheres have been analysed by comparing the Edwards--Slingo radiation scheme to discrete ordinate line-by-line calculations.

The ES radiation scheme's performance in these tests shows that we have successfully adapted it to hot Jupiter-like atmospheres. Our main conclusions are:
\begin{itemize}
\item
Pressure broadening parameters for high-temperature molecular lines are very uncertain and usually extrapolated from room temperature and pressure and small quantum numbers. Improvements in this area will become important as higher accuracy will be required to analyse results from future exoplanet characterisation projects (e.g. JWST, EChO, Sphere and ELT). 
%
\item
A diffusivity factor of $D = 1.66$, already widely used in both Earth and hot Jupiter GCMs, yields the smallest errors from the two-stream approximation, although $D = \sqrt{3} \approx 1.73$ is only slightly less accurate.
\item
About $10$ $k$-coefficients in each band for molecular line absorption yield satisfactory accuracy. Using $\sim 100$ $k$-coefficients per band does improve the overall accuracy, but errors decrease by less than $\SI{50}{\percent}$, while the radiative transfer computation time increases by a factor of $10$. We therefore choose to adopt the former as a balance between accuracy and computational cost.
\item
Both the two-stream approximation and the correlated-$k$ method contribute non-negligibly to the total error, with overall heating rate errors of $\lesssim \SI{10}{\percent}$ in regions with significant heating/cooling. Flux errors are similar or smaller.
\item
Whether a black-body spectrum, solar spectrum or uniform (in wavenumber) weighting scheme is used has little effect on the overall accuracy given the band structure used here (\cref{tbl:bands}). We therefore choose to adopt a uniform weighting scheme, enabling the use of the same $k$-coefficients in both the thermal and stellar spectral regions and for different irradiation spectra.
\item
Using a mean absorption coefficient in each band, as in \citet{Dobbs-Dixon2013}, 
yields inaccurate fluxes and heating rates for molecular absorption. Heating rate errors can reach $\SI{100}{\percent}$ or more, even in regions with significant heating. Band-averaged absorption coefficients should thus be used with caution.
\end{itemize}

Any radiation scheme applied to hot Jupiters should be checked against the tests we have presented here. These tests and the detailed descriptions of our methods and approximations will be useful for future adaptation of radiation schemes in other GCMs. Current observational constraints on exoplanets do not require the level of accuracy we have applied in this work. The field develops at an amazing pace, however, and modellers should now develop the best theoretical and numerical tools to tackle the challenges posed by the increasing accuracy expected from future large observational projects.

\begin{acknowledgements}
We would like to thank Derek Homeier, Jonathan Tennyson, Bernd Freytag, Bertrand Plez, France Allard and Travis Barman for insightful discussions. This work is supported by the European Research Council under the European Community's Seventh Framework Programme (FP7/2007-2013 Grant Agreement No. 247060) and by the Consolidated STFC grant ST/J001627/1. This work is also partly supported by the Royal Society award WM090065. The calculations for this paper were performed on the DiRAC Facility jointly funded by STFC, the Large Facilities Capital Fund of BIS, and the University of Exeter.
\end{acknowledgements}

\appendix

\section{Test 0} \label{ap:test0}

This test is based on a grey atmosphere without scattering and irradiation at the top of the atmosphere, and a lower boundary that emits as a perfect black body at a temperature $T_\text{c}$. These assumptions are consistent with the thermal component of the radiation. To facilitate analytical treatment, we make an additional assumption: the lower boundary is located at a constant optical depth $\tau = \tau^*$. This is done in both the ES radiation scheme and Atmo by explicitly keeping the total mass absorption coefficient, $k_\rho$, constant as a function of pressure and placing the lower boundary at a constant pressure.

\subsection{Analytical solutions}

Analytical solutions of the two-stream approximated and full radiative transfer equation are available under these specific conditions and are provided below. The analytical solutions are compared to the numerical solutions obtained by the ES radiation scheme and to the discrete ordinate solution from Atmo.

\subsubsection{The two-stream approximation}

The two-stream approximated radiative transfer equation in the thermal region, ignoring scattering, is given by \cref{eq:RT_TS}. We now drop the diffuse flux subscript since stellar irradiation is ignored. Using \cref{eq:tau} and assuming hydrostatic equilibrium, the optical depth can be related to the pressure by
\begin{equation}
\tau(\nu, P) = \frac{1}{g} \int_0^P \md P' \, k_\rho(\nu, P') = \frac{k_\rho}{g} P,
\end{equation}
since $k_\rho(\nu, P) = k_\rho$ is assumed to be independent of both frequency $\nu$ and pressure $P$. The optical depth is therefore proportional to pressure and substitution can be done using the equation above.

Integrating \cref{eq:RT_TS} with respect to frequency yields
\begin{equation}
\pm \frac{1}{D} \frac{\md F^\pm (\tau)}{\md \tau} = F^\pm (\tau) - \sigma T_\text{c}^4,
\end{equation}
where Stefan-Boltzmann's law has been used. The above equation is a simple inhomogeneous linear first order differential equation in optical depth, $\tau$, and can be solved using traditional techniques. The homogeneous solution, i.e. ignoring the Planck emission, is given by
\begin{equation}
F_\text{h}^\pm (\tau) = A_\pm e^{\pm D \tau},
\end{equation}
where $A_\pm$ is determined by boundary conditions, while the particular solution in this case is given by
\begin{equation}
F_\text{p}^\pm (\tau) = \sigma T_\text{c}^4,
\end{equation}
which yields the complete solution
\begin{equation}
F^\pm (\tau) = F_\text{h}^\pm (\tau) + F_\text{p}^\pm (\tau) = A_\pm e^{\pm D \tau} + \sigma T_\text{c}^4 .
\end{equation}
At the upper boundary, i.e. $\tau = 0$, we have
\begin{equation}
F^- (\tau = 0) = A_- + \sigma T_\text{c}^4 = 0 \quad \Rightarrow \quad A_- = - \sigma T_\text{c}^4.
\end{equation}
At the lower boundary, which we place at an optical depth of $\tau = \tau^*$, we have
\begin{equation}
F^+ (\tau = \tau^*) = A_+ e^{D \tau^*} + \sigma T_\text{c}^4 = \sigma T_\text{c}^4 \quad \Rightarrow \quad A_+ = 0.
\end{equation}
The upwelling, downwelling and total fluxes are therefore
\begin{align}
F^+ (\tau) &= \sigma T_\text{c}^4, \\
F^- (\tau) &= \sigma T_\text{c}^4 \left[ 1 - e^{- D \tau} \right], \\
F(\tau) &= F^+ (\tau) - F^- (\tau) = \sigma T_\text{c}^4 e^{-D \tau} .
\label{eq:F_TS_E1}
\end{align}
The heating rate is given by \cref{eq:heating_rate}, and using \cref{eq:dtau}, we get
\begin{align}
\mathcal H &= -\frac{\md F}{\md z} = k_\rho \rho \frac{\md F}{\md \tau} = \frac{k_\rho P \bar m}{R T} \frac{\md F}{\md \tau} \\
&= - \frac{k_\rho P \bar m D}{R T} \sigma T_\text{c}^4 e^{- D \tau} \notag \\
&= - \frac{k_\rho P \bar m D}{R T} \sigma T_\text{c}^4 e^{- D k_\rho P/g} .
\label{eq:H_TS_E1}
\end{align}

\subsubsection{The angularly dependent radiative transfer equation}

The full angular dependent (but still azimuthally averaged) radiative transfer equation without scattering is given by \citep{Thomas2002}
\begin{equation}
u \frac{\md I_\nut (\tau, u)}{\md \tau} = I_\nut (\tau, u) - B_\nut (T_\text{c}) ,
\end{equation}
where $u = \cos \theta$. From above, the general solution is given by
\begin{equation}
I_\nut (\tau, u) = A_\nut(u)e^{\tau/u} + B_\nut(T_\text{c}) .
\end{equation}
Note that a discrete ordinate method gives the same equation and solution, except $u$ is replaced by the quadrature points $u_i$. No downward radiation at the upper boundary implies $I_\nut(\tau = 0, u < 0) = 0$, which yields
\begin{equation}
I_\nut(\tau, u < 0) = B_\nut(T_\text{c}) \left[ 1 - e^{\tau/u} \right].
\label{eq:I_u<0_E1}
\end{equation}
Perfect black-body radiation in the upward direction at the lower boundary implies $I_\nut(\tau = \tau^*, u > 0) = B_\nut(T_\text{c})$, which yields
\begin{equation}
I_\nut(\tau, u > 0) = B_\nut(T_\text{c}) .
\label{eq:I_u>0_E1}
\end{equation}
Note that $I_\nut(\tau, u)$ is anisotropic in the downward direction except in the limit of high optical depths, $\tau \to \infty$. The intensity at a given optical depth in the downward direction, $u < 0$, is dictated by the amount of atmosphere above it, in the direction of the radiation, emitting thermally. The ``effective optical depth'' or optical path is higher for smaller values of $\bar \mu$, and the intensity consequently cannot be isotropic. The exception is at high optical depths, where the atmosphere becomes optically thick in all directions.

The upward flux is
\begin{equation}
F_\nut^+ (\tau) = 2\pi \int_0^1 \md \mu \, \mu I_\nut(\tau, \mu)
= 2\pi B_\nut(T_\text{c}) \int_0^1 \md \mu \, \mu
= \pi B_\nut(T_\text{c}),
\end{equation}
while the downward flux is
\begin{align}
F_\nut^- (\tau) &= 2\pi \int_0^1 \md \mu \, \mu I_\nut(\tau, -\mu)
= 2\pi B_\nut(T_\text{c}) \int_0^1 \md \mu \, \mu \left[ 1 - e^{-\tau/\mu} \right] \\
&= \pi B_\nut(T_\text{c}) - 2\pi B(T_\text{c}) \int_0^1 \md \mu \, \mu e^{-\tau/\mu} .
\end{align}
This integral does not have a simple closed-form solution, but can be found numerically. Integrating over all wavenumbers and using Stefan-Boltzmann's law, the total flux is given by
\begin{equation}
F(\tau) = F^+ (\tau) - F^- (\tau) = 2\sigma T_\text{c}^4 \int_0^1 \md \mu \, \mu e^{-\tau/\mu},
\label{eq:F_FULL_E1}
\end{equation}
i.e. dictated by this integral and clearly not equivalent to \cref{eq:F_TS_E1}. Note that the two-stream approximation effectively evaluates the integral in \cref{eq:F_FULL_E1} using a single quaderature point $\bar \mu = 1/D$ which can be chosen using e.g. Gauss--Legendre quadrature or an empirical fit. The heating rate, \cref{eq:heating_rate}, is similarly given by
\begin{align}
\mathcal H &= \frac{k_\rho P \bar m}{RT} \frac{\md F}{\md \tau} = \frac{2 k_\rho P \bar m}{RT} \sigma T_\text{c}^4 \int_0^1 \md \mu \, \mu \frac{\md}{\md \tau} \left[ e^{-\tau/\mu} \right] \\
&= -\frac{2 k_\rho P \bar m}{RT} \sigma T_\text{c}^4 \int_0^1 \md \mu \, e^{-\tau/\mu} .
\label{eq:H_FULL_E1}
\end{align}

\subsection{Accuracy of the two-stream approximation}

\begin{figure}
\includegraphics[width=\columnwidth]{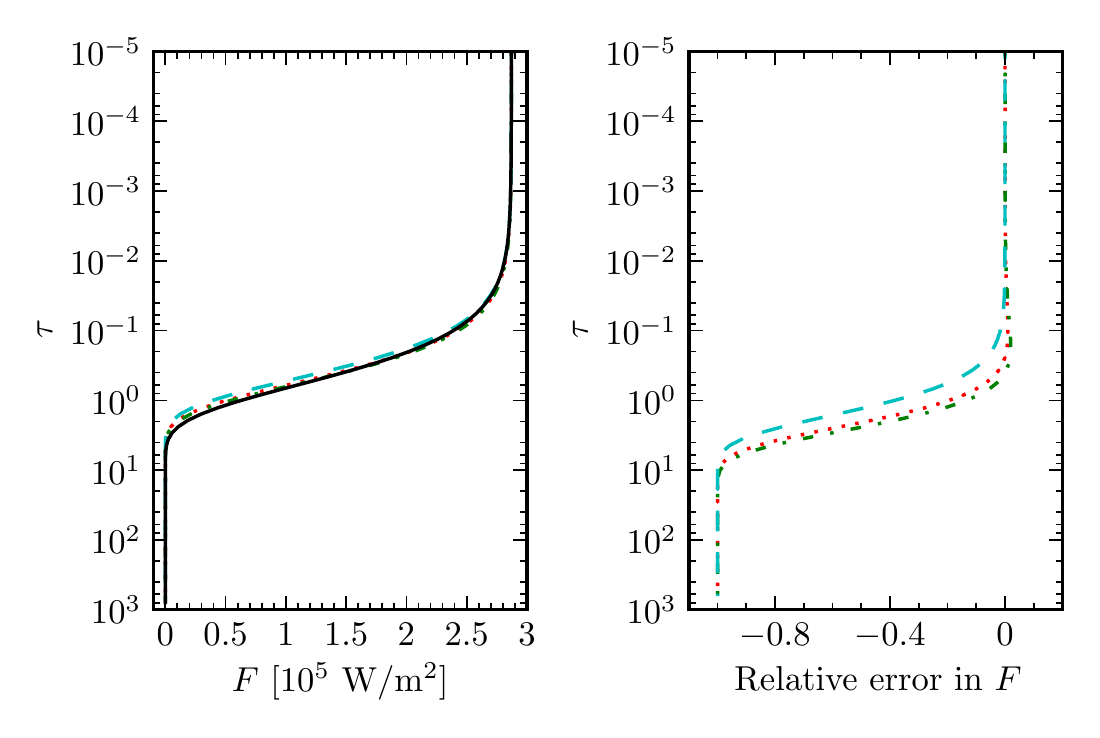}
\caption{The left-hand panel shows the fluxes obtained using the two-stream flux in \cref{eq:F_TS_E1} and exact flux in \cref{eq:F_FULL_E1} obtained with $D = \sqrt{3}$ (dotted, red), $D = 1.66$ (dashed-dotted, green), $D = 2$ (dashed, cyan) and solving the fully angular dependent radiative transfer equation (solid, black). The right-hand panel shows the calculated relative errors in the two-stream fluxes. Relative errors become unreasonably large only where the flux is very small.}
\label{fig:test0_nflx}
\end{figure}

\begin{figure}
\includegraphics[width=\columnwidth]{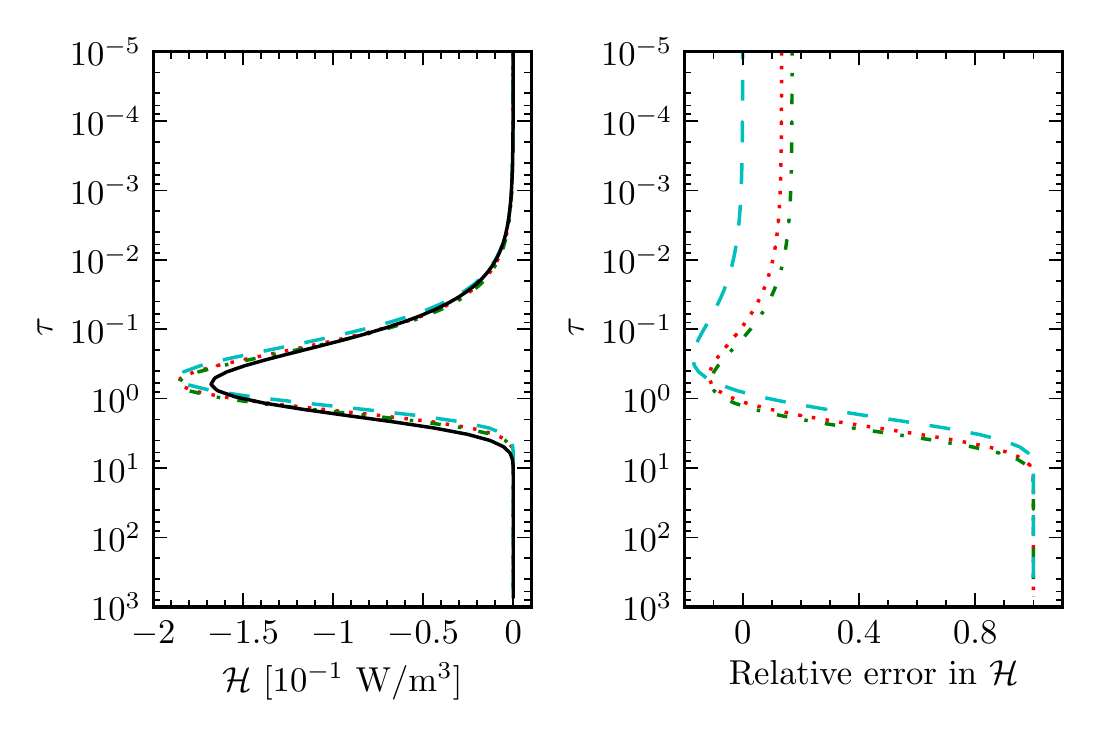}
\caption{Same as \cref{fig:test0_nflx} for heating rates. Relative errors become unreasonably large only where the heating rate is very small.}
\label{fig:test0_hrts}
\end{figure}

Fluxes and heating rates with errors are plotted in \cref{fig:test0_nflx,fig:test0_hrts} using the solutions in \cref{eq:F_TS_E1,eq:H_TS_E1,eq:F_FULL_E1,eq:H_FULL_E1}. At small optical depths, the flux is equal to the black-body flux while at large optical depths, the flux is zero, as expected. The heating rate is zero at both low and high optical depths, while at intermediate optical depths the atmosphere is cooled (the heating rate is negative). Interestingly, the relative error in both flux and heating rate approaches unity at large optical depths, a consequence of the two-stream solutions approaching zero faster than the full solution. We do not consider this a problem, however, since both the flux and heating rate are close to zero in this region.

In \cref{tbl:test0_ts} we show the calculated $L^1$ norms using the analytical solutions derived above. The smallest errors are achieved with $D=1.66$, but using $D = \sqrt{3}$ only yields slightly larger errors. This is verified by looking at the relative error in \cref{fig:test0_nflx,fig:test0_hrts}.

\begin{table}
\centering
\caption{Computed flux ($F$) and heating rate ($\mathcal H$) $L^1$ norms for test~0 using the analytical solutions, thereby eliminating the errors from the numerical solution schemes. The smallest errors are obtained with $D = 1.66$.}
\begin{tabular}{l|r|r}
 & $L^1$, $F$ & $L^1$, $\mathcal H$ \\ \hline
$D = \sqrt{3}$ & $0.007$ & $0.103$ \\
$D = 1.66$ & $0.006$ & $0.094$ \\
$D = 2$ & $0.015$ & $0.174$
\end{tabular}
\label{tbl:test0_ts}
\end{table}

It is worth noting that the different values for $D$ yield different convergence towards zero heating rate at low optical depths, evident in the right-hand panel of \cref{fig:test0_hrts}, caused by the factor $D$ in \cref{eq:H_TS_E1}. Comparing \cref{eq:H_TS_E1,eq:H_FULL_E1}, it is clear that only $D = 2$ will yield the correct behaviour of the heating rate at low optical depths. The effect on the heating rate itself is small, however, and $D = 1.66$ yields the most correct heating rate overall.

\subsection{Accuracy of the numerical scheme}

\begin{figure}
\includegraphics[width=\columnwidth]{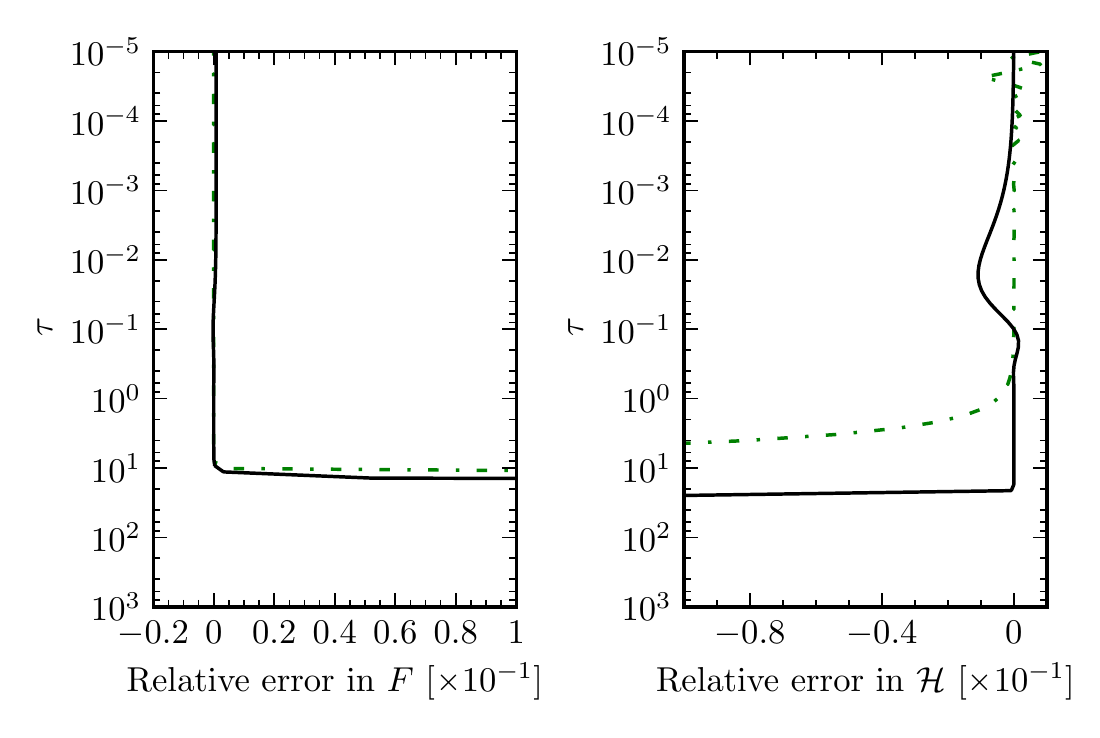}
\caption{Relative error in the numerical solutions from the ES radiation scheme (dashed-dotted, green) and Atmo (solid, black) calculated using the analytical solution in \cref{eq:F_TS_E1,eq:H_TS_E1,eq:F_FULL_E1,eq:H_FULL_E1}. Relative numerical errors become large at large optical depths, caused by both fluxes and heating rates being very small.}
\label{fig:test0_num_error}
\end{figure}

We use the analytical expressions in~\cref{eq:F_TS_E1,eq:H_TS_E1,eq:F_FULL_E1,eq:H_FULL_E1} to estimate the errors in the numerical solution schemes in both the ES radiation scheme and Atmo. We have plotted the numerical error in \cref{fig:test0_num_error} as a function of optical depth, and given the $L^1$ errors in \cref{tbl:test0_num}. The errors are small for small optical depths, while at large optical depths the errors increase significantly. The error in the flux and heating rate reach $\SI{10}{\percent}$ at about $\tau = 10$ and $\tau = 4$, respectively, for the ES radiation scheme, while Atmo is accurate to a somewhat larger optical depth. The $L^1$ errors reflects this, keeping in mind that the error in Atmo is also caused by a finite number of rays in the Gaussian quadrature, which may become important at the accuracy level of the numerical solver. Both numerical schemes are seen to yield errors significantly smaller than errors caused by the two-stream approximation. This confirms that both numerical solvers yield satisfactory accuracy.

\begin{table}
\centering
\caption{Computed flux ($F$) and heating rate ($\mathcal H$) $L^1$ norms for test~0 comparing the numerical and analytical solutions to check the accuracy of the numerical schemes.}
\begin{tabular}{l|r|r}
 & $L^1$, $F$ & $L^1$, $\mathcal H$ \\ \hline
ES radiation scheme & $\num{3.88e-5}$ & $\num{3.55e-3}$ \\
Atmo & $\num{6.20e-4}$ & $\num{1.22e-3}$
\end{tabular}
\label{tbl:test0_num}
\end{table}

\bibliographystyle{aa}
\bibliography{sources}

\end{document}